\documentclass[3p,floatfix,showpacs,tightenlines,showkeys,superscriptaddress,amsmath,amssymb,nofootinbib]{elsarticle}

\usepackage{amssymb,amsbsy,epsfig,color,graphicx}
\usepackage{color}
\usepackage{subfigure}
\usepackage{[longtable}
\usepackage{array}
\usepackage{dcolumn}   
\usepackage{cellspace}
\usepackage{mathtools}
\usepackage{amstext}
\usepackage{amssymb}
\usepackage{stmaryrd}
\usepackage{stackrel}
\usepackage{graphicx}
\usepackage{esint}
\usepackage[utf8]{inputenc}
\usepackage{blindtext}
\usepackage{float}
\restylefloat{table}
\usepackage{booktabs}
\usepackage{enumitem}

\usepackage{etoolbox} 
\usepackage{lipsum} 
\usepackage[capitalize]{cleveref}

\journal{Nuclear Physics B}

\makeatletter

\appto{\appendix}{%
	\@ifstar{\def\theequation@prefix{A.}}%
	{}%
}
\makeatother

\begin{document}

\begin{frontmatter}

\title{Ghost-free infinite derivative quantum field theory}

\author[first,second,third]{Luca Buoninfante}
\ead{lbuoninfante@sa.infn.it}

\author[first,second]{Gaetano Lambiase}
\ead{lambiase@sa.infn.it}

\author[third]{Anupam Mazumdar}

\ead{anupam.mazumdar@rug.nl}

\address[first]{Dipartimento di Fisica "E.R. Caianiello", Universit\`a di Salerno, I-84084 Fisciano (SA), Italy}
\address[second]{INFN - Sezione di Napoli, Gruppo collegato di Salerno, I-84084 Fisciano (SA), Italy}
\address[third]{Van Swinderen Institute, University of Groningen, 9747 AG, Groningen, The Netherlands}

\begin{abstract}
In this paper we will study Lorentz-invariant, infinite derivative quantum field theories, where infinite derivatives give rise to non-local interactions at the energy scale $M_s$, beyond the Standard Model. We will study a specific class, where there are no {\it new dynamical degrees of freedom} other than the original ones of the corresponding local theory. We will show that the Green functions are modified by a non-local extra term that is responsible for acausal effects, which are confined in the region of non-locality, i.e. $M_s^{-1}.$  The standard time-ordered structure of the causal Feynman propagator is not preserved and the non-local analog of the retarded Green function turns out to be non-vanishing for space-like separations. As a consequence the local commutativity is violated. Formulating such theories in the non-local region with Minkowski signature is not sensible, but they have Euclidean interpretation. We will show how such non-local construction ameliorates ultraviolet/short-distance singularities suffered typically in the local quantum field theory. We will show that non-locality and acausality are inherently off-shell in nature, and only quantum amplitudes are physically meaningful, so that all the perturbative quantum corrections have to be consistently taken into account.
\end{abstract}


\end{frontmatter}


\section{Introduction and beyond 2 derivatives}

In nature, a simple 2 derivative field theory is able to capture aspects of local interactions -  both in a classical and in a quantum sense. However neither locally nor globally, nature forbids going beyond 2 derivative kinetic terms. In this sense, there is no prohibition in constructing higher derivative Lorentz-invariant (and diffeomorphism invariant in the context of curved spacetime) kinetic terms. Higher derivative kinetic terms {\it may } harbor certain kind of classical and quantum instabilities depending on the nature of the sign of the kinetic terms; for example, Ostr\'ogradsky instability~\cite{ostrog} can arise, due to the fact that the Hamiltonian density is unbounded from below. This classical instability can also be seen at a quantum level, in the Lagrangian formalism, especially when there are extra propagating degrees of freedom, which comes with a negative residue in the propagator- an indication of a ghost-like degree of freedom. Typically, such instabilities are considered to be safe in low energy effective field theories, at energy scales much below the cut-off, but the {\it ghost} problem becomes important at high energies, towards the ultraviolet (UV) scales. There is one particular avenue, where higher derivatives play a very significant role - which is massless gravitational interaction. 

It has been known for a while that the quadratic curvature theory of gravity is renormalizable in $4$ dimensions~\cite{stelle}~\footnote{Quadratic curvature action contains terms like ${\cal R}^2,~{\cal R}_{\mu\nu}{\cal R}^{\mu\nu},~{\cal C}^{\mu\nu\rho\sigma}{\cal C}_{\mu\nu\rho\sigma}$, where $\mu, \nu=0,1,2,3$, $\mathcal{C}$ stands for the Weyl tensor. In $4$ dimensions one can further reduce the action with the help of Gauss-Bonnet identity.}, but contains a massive spin-2 Weyl {\it ghost} as a dynamical degree of freedom. Indeed, being a $4$-dimensional higher derivative theory of gravity, it improves the UV behavior of gravitational interaction, but not sufficiently strong enough to resolve some of the thorny issues of gravity - such as classical singularity problems in cosmology and blackhole solutions; these singular solutions still persist. Recently, it has been noticed that theories with kinetic terms made of derivatives of infinite order are better equipped to handle the issue of {\it ghost}. In fact, this classic observation was made in the context of gravity and gauge theory first~\cite{krasnikov,kuzmin,Tomboulis:1980bs,Tomboulis:1997gg}.

In particular, in Ref.\cite{Biswas:2011ar} it was explicitly shown that the most general quadratic curvature gravitational action (parity-invariant and torsion-free), with infinite covariant derivatives can make the gravitational sector free from the Weyl {\it ghost} and, moreover, the infinite derivative action is free from classical singularities, such as blackhole type~\cite{Biswas:2011ar,Biswas:2013cha,Frolov:2015bia,Frolov:2015usa,Koshelev:2018hpt,Koshelev:2017bxd,Buoninfante:2018xiw,Cornell:2017irh,Buoninfante:2018rlq,Buoninfante:2018stt}~\footnote{Previously, arguments were provided regarding non-singular solutions in Refs.~\cite{Tseytlin:1995uq,Siegel:2003vt}.} and cosmological type~\cite{Biswas:2005qr,Biswas:2006bs,Biswas:2010zk,Biswas:2012bp,Koshelev:2012qn,Koshelev:2017tvv,Koshelev:2018rau}. The modified graviton propagator around the Minkowski background in $4$ dimensions is given by \cite{Biswas:2011ar}
\begin{equation}
\Pi(-k^2)=\frac{1}{a(-k^2)}\Pi_{\scriptscriptstyle GR}(-k^2), \label{eq:2}
\end{equation}
where $\Pi_{\scriptscriptstyle GR}(-k^2)=\mathcal{P}^2/k^2-\mathcal{P}_s^0/2k^2$ is the graviton propagator in Einstein's general relativity (GR) expressed in terms of the spin-projection operators along the spin-$2$ and spin-$0$ components, respectively~\footnote{See Refs. \cite{Biswas:2011ar,Biswas:2013kla,Buoninfante:2016iuf} for a pedagogical review on the spin-projection formalism and its application to the computation of the graviton propagator.}. The presence of infinite covariant derivatives are captured by $a(-k^2)$, which can contain in principle {\it infinitely} many poles. This means {\it infinitely} many new degrees of freedom, other than the massless graviton propagating in $4$ dimensions. The key observation here is to {\it avoid} the presence of the extra degrees of freedom, and keep {\it solely} the original transverse and traceless graviton as the only dynamical degree of freedom. In order to avoid {\it extra} poles in the propagator the form of $a(-k^2)$ is constrained by~\cite{Tomboulis:1980bs,Tomboulis:1997gg,Biswas:2005qr,Biswas:2011ar} :
\begin{equation}\label{prop}
a(-k^2)= e^{\gamma(k^2/M_s^2)},~~~~~~a(-k^2)\rightarrow 1\,\,\,\,{\rm if}\,\,\,\,k/M_s\rightarrow 0,
\end{equation}
where $\gamma(k^2)$ is an {\it entire function} which ensures that there are no extra poles in the complex plane and $M_s$ is the new scale of physics~\footnote{It is worth mentioning that such a scale of non-locality $M_s$ has been constrained in different  field theories. For instance, in the case of  gravitational interaction one has the lower bound $M_s>0.004$eV coming from torsion balance experiments as pointed out in Ref.\cite{Edholm:2016hbt}.}. The absence of {\it ghosts} can be understood by the fact that there are no new dynamical degrees of freedom left in the propagator\footnote{See also Ref.\cite{Becker:2017tcx} where the ghost problem has been discussed in the non-perturbative scenario of asymptotic safety.}. At low energies $k\ll M_s$, the quadratic curvature graviton propagator in Eq. \eqref{eq:2} reduces to that of the Einstein-Hilbert propagator of GR in $4$ dimensions, as expected; while at high energies, $k\gg M_s$, the graviton propagator is exponentially suppressed. Gravitational interaction is derivative in nature, therefore the vertex operator gets modified by an exponential enhancement. The interplay between the graviton propagator and the vertex operator leads to the non-locality in the momentum space. The structure of non-locality is hidden in the form factor $a(-k^2)$, as shown in Ref.~\cite{Biswas:2011ar}. 

Besides having very interesting applications in resolving singularities in blackhole physics and in cosmology, at a quantum level, it is believed that the introduction of such form-factors can make the gravitational theory UV-finite, beyond $1$-loop, as discussed into details in Refs.\cite{kuzmin,Tomboulis:1997gg,Modesto,Biswas:2012ka,Talaganis:2014ida}. In this respect, non-local interactions ameliorate the UV aspects of gravity at short distances and small time scales. Moreover, important progress have been made also in the context of non-local thermal field theory, see Refs.\cite{Biswas:2012ka,Biswas:2009nx,Biswas:2010xq,Biswas:2010yx}.

The appearance of non-locality in string theory is very well known, the infinite derivative operators appear in the string field theory (SFT)~\cite{witten,ohmori,eliezer}, where they are known as $\alpha'$ corrections. In STF  vertices arise of the following form:
\begin{equation}
V\sim e^{c \alpha'\Box} \label{eq:1}
\end{equation}
where $c\sim \mathcal{O}(1)$ is a dimensionless constant that can change depending on whether one considers either open or closed string, and $\alpha'$ is the so called universal Regge slope, and $\Box=\eta_{\mu\nu}\partial^{\mu}\partial^{\nu}$ is the d'Alembertain operator in flat spacetime, where $\eta_{\mu\nu}={\rm diag}(-1,+1,+1,+1)$. Note that $\alpha'=1/M_s^2$ is a dimensionful coupling, where $M_s$ is denoted to be the string tension. At the phenomenological level, there have been attempts to construct a model with infinite derivative Higgs and fermion sector, which indeed ameliorates the UV aspects of the Abelian Higgs~\cite{Biswas:2014yia,Ghoshal:2017egr}. Moreover, in Ref.\cite{Buoninfante:2018gce} it was found that the scale of non-locality $M_s$ is not fixed but is a dynamical quantity, indeed it can shifts towards the infrared regime as a function of the number of particles taking part in the process, meaning that the space-time region on which the non-local interaction happens can become larger as the number of particles increases.

Motivated by the success of the infinite derivative gravity, and the success of open SFT, it is worthwhile to investigate some quantum aspects of infinite derivative field theories in more detail~\footnote{In general, non-locality can be thought {\it at least} in two different ways: (i) as discretization of the space-time;  (ii) or purely related to the interaction in systems defined in a continuum space-time. In the case (i) there would be a minimal length-scale given by the size of the unit-cell in such a discrete background, and it is often identified with the Planck length, $\ell_p\sim 1/M_p$, where $M_p\sim 10^{19}$~GeV, or the string scale below the Planck scale in $4$ dimensions. As for (ii), the non-locality does not affect the kinematics at the level of free-theory, but it becomes relevant only when dynamics is considered. In the free-theory such a non-locality would not play any role, but it would become relevant as soon as the interaction is switched on. In this regard, we will be investigating the latter scenario, where we will consider a continuum space-time and introduce non-locality through  form-factors into either the kinetic operator or the interaction vertex. First attempts along (ii) trace back in the fifties, when people were still facing the problem of ultraviolet (UV) divergences in quantum field theory and renormalization was still not very well understood, thus an alternative possibility to deal with divergences was the introduction of non-local interactions with the aim to regularize the theory and make it finite in the UV. These developments also encouraged a deeper understanding of  field theories from an axiomatic point of view \cite{efimov,pais}.}. We wish to study some properties of Lorentz-invariant infinite derivative quantum field theories with {\it exponential analytic} form-factors made of derivatives of infinite order. We will treat the simplest case of a scalar field. The paper is organized as follows. 

In Section \ref{IDA}, we will introduce the  action for a real scalar field and analyze into details the structure of the propagator, and emphasize that non-locality is important only when the interactions are switched on. We will see how to perform calculations with operators involving derivatives of infinite orders. In Section \ref{CV}, we will show that non-locality leads to a violation of causality in a space-time region whose size is given by the scale of non-locality $l_s=1/M_s.$ We will show that the retarded Green function becomes acausal due to non-locality and as a consequence we show that also local commutativity is violated. In Section \ref{ME}, we will discuss the Euclidean prescription for computing correlators and amplitudes. We will compute the Euclidean $2$-point correlation function and show that it is non-singular at the Euclidean origin. In Section \ref{scattering-amplitudes}, we will discuss quantum scattering amplitudes. In Section \ref{conlus}, we will present summary and conclusions.


\section{Infinite derivative action}\label{IDA}

We now wish to introduce a Lorentz-invariant infinite derivative field theory for a real scalar field $\phi(x)$ by an action: 
\begin{equation}
S=\frac{1}{2}\int d^4x d^4y\phi(x)\mathcal{K}(x-y)\phi(y) -\int d^4x V(\phi(x)), \label{eq:5}
\end{equation}
where the operator $\mathcal{K}(x-y)$ in the kinetic term makes explicit the dependence on the field variables at finite distances $x-y$, signaling the presence of a non-local nature; the second contribution to the action is a standard local potential term. We can rewrite the kinetic term as follows
\begin{equation}
\begin{array}{rl}
S_{\scriptscriptstyle K}= & \displaystyle \frac{1}{2}\int d^4x d^4y\phi(x)\mathcal{K}(x-y)\phi(y)\\
= & \displaystyle \frac{1}{2}\int d^4x d^4y\phi(x)\int \frac{d^4k}{(2\pi)^4}F(-k^2)e^{ik\cdot(x-y)}\phi(y)\\
= & \displaystyle \frac{1}{2}\int d^4x d^4y\phi(x)F(\Box) \int\frac{d^4k}{(2\pi)^4}e^{ik\cdot(x-y)}\phi(y)\\
= & \displaystyle \frac{1}{2}\int d^4x\phi(x)F(\Box)\phi(x),
\end{array} \label{eq:6}
\end{equation}
where $F(-k^2)$ is the Fourier transform of $\mathcal{K}(x-y)$, and we have used the integral representation of the Dirac delta, $\int\frac{d^4k}{(2\pi)^4}e^{ik\cdot(x-y)}=\delta^{(4)}(x-y)$. From Eq.\eqref{eq:6} note that the  operator $\mathcal{K}(x-y)$ has the following general form \cite{tomboulis2015}:
\begin{equation}
\mathcal{K}(x-y)=F(\Box)\delta^{(4)}(x-y). \label{eq:7}
\end{equation}
Note that the action in Eqs.\eqref{eq:5}-\eqref{eq:6} is manifestly {\it Lorentz invariant}, thus it is possible to define a divergenceless stress-energy momentum tensor \cite{barci}. 
Note that $\Box$ is dimensionful, and strictly speaking we should write $\Box/M_s^2$. For brevity, we will suppress $M_s$ in the definition of the form factors from now on. Further note that the action without the potential has {\it no} non-locality. The homogeneous solution obeys the local equations of motion, i.e. 
the plane wave solution of the local field theory, see discussion in Section \ref{FRNLI}.

\subsection{Choice of kinetic form factor}\label{KFF}

So far we have not required any property for the form factor $F(\Box)$\footnote{In the following we will not refer to the operator $\mathcal{K}(x-y)$ anymore, but we will speak in terms of $F(\Box)$.}, other than being Lorentz invariant; however it has to satisfy special conditions in order to define a consistent quantum field theory, in particular absence of {\it ghosts} at the tree level. We will restrict the class of operators by demanding $F(\Box)$ to be an {\it entire analytic function}\footnote{Let us remind that an entire function is a complex-valued function that is holomorphic at all finite points in the whole complex plane. It is worthwhile to mention that in literature there are also examples of field theory where the operator is a non-analytic function. For instance, from quantum correction to the effective action of quantum gravity non-analytic terms like $\mathcal{R}(\mu^2/\Box)\mathcal{R}$ and $\mathcal{R}{\rm ln}(\Box/\mu^2)\mathcal{R}$ emerge \cite{Bravinsky,Deser:2007jk,Belgacem:2017cqo,Woodard:2018gfj}. Moreover, in causal-set theory \cite{sorkin,belenchia}, the Klein-Gordon operator for a massive scalar field is modified as follows
	\begin{equation}
	F(\Box +m^2)= \Box+m^2-\frac{3l_p^2}{2\pi \sqrt{6}}(\Box+m^2)^2\left[3\gamma-2+{\rm ln}\left(\frac{3l_p^2(\Box+m^2)^2}{2\pi}\right)\right]+\cdots, \label{eq:8}
	\end{equation}
	where $\gamma$ is the Euler-Mascheroni constant and $\ell_p$ is the appropriate length scale; note also the presence of branch cuts once analyticity is given up.}. We can now apply the Weierstrass factorization-theorem for entire functions, so that we can write:
\begin{equation}
F(\Box)=e^{-f(\Box)}\prod\limits_{i=1}^{N}(\Box-m_i^2),\label{eq:9}
\end{equation}
where $f(\Box)$ is also an entire function, $N$ can be either finite or infinite and it is related to the number of zeros of the entire function $F(\Box)$. From a physical point of view, $2N$ counts the number of poles in the propagator that is defined as the inverse of the kinetic operator in Eq.\eqref{eq:9}. The exponential function does not introduce any extra degrees of freedom and it is suggestive of a cut-off factor that could improve the UV-behavior of loop-integrals in perturbation theory, moreover it contains all information about the infinite-order derivatives:
\begin{equation}
e^{-f(\Box)}=\sum\limits_{n=0}^{\infty}\frac{f_n}{n!}\Box^n,\label{eq:10}
\end{equation}
where $f_n:=\left.\partial^{(n)} e^{-f(\Box)}/\partial \Box^n\right|_{\Box=0}$.
By inverting the kinetic operator in Eq. \eqref{eq:9}, we obtain the propagator that in momentum space reads~\footnote{We adopt the convention in which the propagator in the Minkowski signature is defined as the inverse of the kinetic term times the imaginary number $'i'$.}
\begin{equation}
\Pi(k)=e^{f(-k^2)}\prod\limits_{i=1}^{N}\frac{-i}{k^2+m_i^2}.\label{eq:11}
\end{equation}
One can immediately notice that if $N>1$ ghosts appear. Indeed, we can decompose the propagator in Eq.\eqref{eq:11} as
\begin{equation}
e^{f(-k^2)}\prod\limits_{i=1}^{N}\frac{1}{k^2+m_i^2}=e^{f(-k^2)}\sum\limits_{i=1}^{N}\frac{c_i}{k^2+m_i^2},\label{eq:12}
\end{equation}
where the coefficients $c_i$ contain the sign of the residues of the propagator at each pole; then by multiplying with $e^{-f(-k^2)}k^2$, and taking the limit $k^2\rightarrow\infty$, we obtain
\begin{equation}
0=\sum\limits_{i=1}^N c_i, \label{eq:13}
\end{equation}
which means that at least one of the coefficients $c_i$ must be negative in order to satisfy the equality in Eq.\eqref{eq:13}, i.e., at least one of the degrees of freedom must be {\it ghost} like. In this paper we will focus on the case $N=1$, so that tree-level unitarity will be preserved and {\it no ghosts} whatsoever will be present in the physical spectrum of the theory.

Let us now fix the function $f(\Box)$ in the exponential. As we have already mentioned, it has to be an entire function, moreover it has to recover the local Klein-Gordon operator, i.e. $2$-derivatives differential operator, in the IR regime, $\Box/M_s^2 \rightarrow 0$. In this paper we will mainly consider polynomial functions of $\Box$, in particular we will study the simplest operator~\footnote{See also Ref.\cite{Tomboulis:1997gg,Modesto,Edholm:2016hbt} for other possible choices of entire functions that improve the UV-behavior.}
\begin{equation}
f(\Box)=-\frac{(-\Box+m^2)^n}{M_{s}^{2n}} \Longrightarrow F(\Box)=e^{\frac{(-\Box+m^2)^n}{M_{s}^{2n}}}(\Box-m^2), \label{eq:14}
\end{equation}
where $n$ is a positive integer and we have explicitly reinstated $M_s$. In the infinite derivative gravitational action, the form of $f(\Box)$ remains very similar, except $m=0$ \cite{Biswas:2011ar}.

\subsection{Field redefiniton and  non-local interaction}\label{FRNLI}

The infinite derivative field theory introduced in Eqs.\eqref{eq:5} and \eqref{eq:6} shows a  modification in the kinetic term. However, note that we can also define an infinite derivative field theory where the kinetic operator corresponds to the usual local Klein-Gordon operator by making the following field re-definition:
\begin{equation}
\tilde{\phi}(x)= \displaystyle e^{-\frac{1}{2}f(\Box)}\phi(x)
= \displaystyle \int d^4y \mathcal{F}(x-y)\phi(y),
\label{42}
\end{equation}
where $\mathcal{F}(x-y):=e^{-\frac{1}{2}f(\Box)}\delta^{(4)}(x-y)$; the quantity $\mathcal{F}(x-y)$ is the kernel of the differential operator $e^{-\frac{1}{2}f(\Box)}$.
By inserting such a field redefinition into the action in Eq.\eqref{eq:5}, we obtain an equivalent action that we can still name by $S$:
\begin{equation}
S=\frac{1}{2}\int d^4x \tilde{\phi}(x)(\Box-m^2)\tilde{\phi}(x) -\int d^4x V\left(e^{\frac{1}{2}f(\Box)}\tilde{\phi}(x)\right).
\label{43}
\end{equation}
From Eq.\eqref{43} it is evident that now the  form-factor $e^{\frac{1}{2}f(\Box)}$ appears in the interaction term and that non-locality only plays a crucial rule when the interaction is switched on as the free-part is just the standard local Klein-Gordon kinetic term. Such a feature of non-locality is relevant only at the level of interaction, this will become more clear below, when we will discuss homogeneous (without interaction-source), and inhomogeneous (with interaction-source) field equations.


\subsection{Homogeneous field equations: Wightman function}\label{WF}

We can now determine the field equation for a free massive scalar field by varying the kinetic action in Eq.\eqref{eq:6} in the case of $N=1$ degree of freedom, see section~\ref{KFF}, and we obtain
\begin{equation}
F(\Box)\phi(x)=0 \Longleftrightarrow e^{-f(\Box)}(\Box-m^2)\phi(x)=0, \label{eq:15}
\end{equation}
that is a homogeneous differential equation of infinite order. One of the first question one needs to ask is how to formulate the Cauchy problem corresponding to Eq.\eqref{eq:15} or, in other words, whether we really need to assign an infinite number of initial conditions in order to find a solution; if this is the case we would lose physical predictability as we would need an infinite amount of information to uniquely specify a physical configuration. Fortunately, as pointed out in Ref.\cite{barnaby,Calcagni:2018lyd}, what really fixes the number of independent solutions is the pole structure of the inverse operator $F^{-1}(\Box)$. For instance, as for Eq.\eqref{eq:15} we have two poles solely given by the Klein-Gordon operator $\Box-m^2$, which implies that the number of initial conditions and independent solutions is also two.

In particular, note that the equality $(\Box-m^2)\phi(x)=0$ also solves Eq.\eqref{eq:15}, namely the two independent solutions of Eq.\eqref{eq:15} are given by the same two independent solutions of the standard local Klein-Gordon equation~\footnote{The normalization factor $\frac{1}{(2\pi)^3\sqrt{2\omega_{\vec{k}}}}$ in the field-decomposition Eq.\eqref{eq:16} is consistent with the following conventions for the creation operator $a^{\dagger}_{\vec{k}}|0\rangle = \frac{1}{\sqrt{2\omega_{\vec{k}}}}|\vec{k}\rangle,$ for the states-product $\langle \vec{k}|\vec{k}' \rangle=2\omega_{\vec{k}}(2\pi)^3\delta^{(3)}(\vec{k}-\vec{k}')$ and for the identity in the Fock space $\mathbb{I}=\int \frac{d^3k}{(2\pi)^3}\frac{1}{2\omega_{\vec{k}}}|\vec{k}\rangle \langle \vec{k}|.$ With such conventions, the canonical commutation relation for free-fields reads $[\phi(x),\pi(y)]_{x^0=y^0}=\delta^{(3)}(\vec{x}-\vec{y}),$ where $\pi(y)$ is the conjugate momentum to $\phi(y).$ }:
\begin{equation}
\phi(x)=\int \frac{d^3k}{(2\pi)^3}\frac{1}{\sqrt{2\omega_{\vec{k}}}}\left(a_{\vec{k}}e^{ik\cdot x}+a^{*}_{\vec{k}}e^{-ik\cdot x}\right), \label{eq:16}
\end{equation}
where $k\cdot x=-\omega_{\vec{k}}x_0+\vec{k}\cdot \vec{x}$, with $\omega_{\vec{k}}=\sqrt{\vec{k}^2+m^2}$. The coefficients $a_{\vec{k}}$ and $a^{*}_{\vec{k}}$ are fixed by the initial conditions and once a quantization procedure is applied they become the usual creation and annihilation operators satisfying the following commutation relations:
\begin{equation}
[a_{\vec{k}},a^{\dagger}_{\vec{k}'}]=(2\pi)^3 \delta^{(3)}(\vec{k}-\vec{k}'),\,\,\,\,\,\,\,\,[a^{\dagger}_{\vec{k}},a^{\dagger}_{\vec{k}'}]=0=[a_{\vec{k}},a_{\vec{k}'}]. \label{eq:17}
\end{equation}
Furthermore, let us remind that the {\it Wightman function} is defined as a solution of the homogeneous differential equation Eq.\eqref{eq:15}, thus from the above considerations it follows that it is not affected by the infinite derivative modification.

Indeed, in a local field theory the Wightman function is found by solving the homogeneous Klein-Gordon equation, and reads~\footnote{Whenever there is a confusion, we will label the local quantities with a subscript  $L$.}
\begin{equation}
W_{\scriptscriptstyle L}(x-y)=\int \frac{d^4k}{(2\pi)^3}\theta(k^0)\delta^{(4)}(k^2+m^2)e^{ik\cdot (x-y)}. \label{eq:18}
\end{equation}
The corresponding infinite derivative Wightman function would be defined by acting on Eq.\eqref{eq:18} with the operator $e^{f(\Box)}$. However, because of the Lorentz-invariance of the operator $e^{f(\Box)}$, with $f(\Box)$ being an entire analytic function,  Eq.\eqref{eq:18} will only depend on $k^2$ in momentum space. Therefore, given the on-shell nature of $W_{\scriptscriptstyle L}(x-y)$ through the presence of $\delta^{(4)}(k^2+m^2)$, one has\footnote{Note that Wightman function for the free-theory can get modified in  field theories with non-analytic form factors, see Refs. \cite{sorkin,belenchia}, in our scenario this is not the case.}
\begin{equation}
W(x-y)= e^{f(\Box)}W_{\scriptscriptstyle L}(x-y)
= \displaystyle e^{f(m^2)} \int \frac{d^4k}{(2\pi)^3}\theta(k^0)\delta^{(4)}(k^2+m^2)e^{ik\cdot (x-y)}. \label{eq:19}
\end{equation}
The exponential operator only modifies the local Wightman function by an overall constant factor $e^{f(m^2)}$ that can be appropriately normalized to 1: $e^{f(m^2)}=1.$ For instance, in the case of exponential of polynomials, as in Eq.\eqref{eq:14}, one has $e^{-(-k^2-m^2)^n/M_s^{2n}}=1,$ once we go on-shell, $k^2=-m^2$. Thus, infinite derivatives {\it do not} modify the Wightman function. It is also clear that the commutation relations between the two free-fields evaluated at two different space-time points will not change:
\begin{equation}
\left\langle0\left|[\phi(x),\phi(y)]\right|0\right\rangle=  W(x-y)-W(y-x)
= W_{\scriptscriptstyle L}(x-y)-W_{\scriptscriptstyle L}(y-x).\label{eq:20}
\end{equation}
Let us remind that for a massive scalar field, one has:
\begin{equation}
\left\langle0\left|[\phi(x),\phi(y)]\right|0\right\rangle= \displaystyle -\frac{i}{2\pi^2}\frac{1}{r} \int\limits_0^{\infty}d|\vec{k}| \frac{|\vec{k}|{\rm sin}(\sqrt{\vec{k}^2+m^2}t){\rm sin}(|\vec{k}|r)}{\sqrt{\vec{k}^2+m^2}}
\equiv  i\Delta (t,r)  
\label{eq:21}
\end{equation}
where we have defined $t=x^0-y^0$ and $\vec{r}=\vec{x}-\vec{y}$; $\Delta(t,r)$ is called Pauli-Jordan function. The above integral  can be calculated, and in the massive case this is given by \cite{bogoliubov}:
\begin{equation}
\Delta(t,r)=-\frac{1}{2\pi}\varepsilon(t)\delta(\rho)+\frac{m}{4\pi\sqrt{\rho}}\theta(\rho)\varepsilon(\rho)J_1(m\sqrt{\rho}),\label{eq:23}
\end{equation}
where $\rho:=t^2-r^2$, $\varepsilon(t) =\theta(t)-\theta(-t)$, and $J_1$ is the Bessel function of the first kind. It is clear that $\Delta(t,r)$ has support only within the past and future lightcones, indeed it vanishes for space-like separations $(\rho<0)$.  When $m=0$,  one has
\begin{equation}
\left.\Delta(t,r)\right|_{m=0}= \displaystyle-\frac{1}{2\pi}\varepsilon(t)\delta(\rho)
= \displaystyle \frac{1}{4\pi r}\left[\delta(t+r)-\delta(t-r) \right],
\label{eq:24}
\end{equation}
which has support {\it only} on the {\it lightcone surface}. By defining the lightcone coordinates $u=t-r$ and $v=t+r$, the massless fields are parametrized by $u=0=v$, as indicated by the Dirac deltas in Eq.\eqref{eq:24}, so it follows that the commutation relations in Eqs.\eqref{eq:23},\eqref{eq:24} define the lightcone structure of the theory, which is not modified by infinite derivatives~\footnote{It is worth mentioning that there are examples of  field theories where the commutation relations for free-fields are modified by the presence of a minimal length-scale. For instance, it happens in non-commutative geometry \cite{douglas} and causal-set theory \cite{sorkin,belenchia}. In the latter the form factor $F(k)$, not only depends on the invariant $k^2$, but also on the sign of $k^0$ signaling the presence of branch-cuts in the Wightman function due to non-analyticity. 
	Furthermore, modified commutation relations may emerge in theories were Lorentz-invariance is broken. Let us consider a very simple pedagogical example where the  form factor explicitly breaks Lorentz-invariance: $F(\nabla^2)=e^{-\nabla^2/M_s^{2}}$, where $\nabla^2\equiv\delta_{ij}\partial^i\partial^j$ is the spatial Laplacian, or in momentum space $F(\vec{k}^2)=e^{\vec{k}^2/M_s^{2}}$. In such a case it is easy to show that the commutator between two free massless scalar fields assumes the following form:
	\begin{equation}
	\left\langle0\left|[\phi(x),\phi(y)]\right|0\right\rangle=  \displaystyle -\frac{i}{2\pi^2}\frac{1}{r} \int\limits_0^{\infty}d|\vec{k}| e^{-\vec{k}^2/M_s^{2}}{\rm sin}(|\vec{k}|t){\rm sin}(|\vec{k}|r)=\displaystyle \frac{iM_s}{8\pi^{3/2}}\left[e^{-\frac{1}{4}M_s^2(r+t)^2}-e^{-\frac{1}{4}M_s^2(r-t)^2}\right].
	\label{eq:25}
	\end{equation}
	It is evident from Eq.\eqref{eq:25} that the commutator for massless free fields is different from zero either inside and outside the lightcone on a region of size $\sim 1/M_s$ around the lightcone surface $u=0=v$.\label{foot1}}.



\subsection{Inhomogeneous field equations: propagator}\label{IFEP}

From the previous considerations it is very clear that non-locality in infinite derivative theories is not relevant at the level of free-theory, but it will play a crucial role when interactions are included. In fact, in presence of the potential term  the field equation is given by
\begin{equation}
e^{-f(\Box)}(\Box-m^2)\phi(x)=\frac{\partial V(\phi)}{\partial \phi(x)}, \label{26}
\end{equation}
and in this case the general solution cannot be simply found by solving the local Klein-Gordon equation, but the exponential operator $e^{-f(\Box)}$ will play a crucial role. Hence, solutions of the inhomogeneous field equation will feel the non-local modification. The simplest example of inhomogeneous equation is the one with a delta source $\delta^{(4)}(x-y)=\delta(x^0-y^0)\delta^{(3)}(\vec{x}-\vec{y})$, whose solution corresponds to the propagator of the theory. In Minkowski signature, the propagator $\Pi(x-y)$ satisfies the following differential equation:
\begin{equation}
e^{-f(\Box_x)}(\Box_x-m^2)\Pi(x-y)=i\delta^{(4)}(x-y), \label{27}
\end{equation}
whose solution can be expressed as
\begin{equation}
\Pi(x-y)=\int \frac{d^4k}{(2\pi)^4}\frac{-ie^{f(-k^2)}}{k^2+m^2-i\epsilon}e^{ik\cdot(x-y)}, \label{28}
\end{equation}
where 
\begin{equation}
\Pi(k)=-\frac{ie^{f(-k^2)}}{k^2+m^2-i\epsilon}, \label{29}
\end{equation}
is the Fourier transform of the propagator in Minkowski signature. We now wish to explicitly show that the propagator in Eq.\eqref{28} can not be identified with the time-ordered product of two fields, $\Pi(x-y)\neq \left\langle 0 \left| T \left( \phi(x)\phi(y)\right) \right| 0 \right\rangle.$
As we have already seen for the Wightman function, the  quantity $\Pi(x-y)$ can be expressed in terms of the local one, $\Pi_{\scriptscriptstyle L}(x-y),$ by acting on the latter with the operator $e^{f(\Box_x)}$:
\begin{equation}
\Pi(x-y)=  \displaystyle e^{f(\Box_x)} \Pi_{\scriptscriptstyle L}(x-y)
=  \displaystyle e^{f(\Box_x)} \left[\theta(x^0-y^0)W_{\scriptscriptstyle L}(x-y)+\theta(y^0-x^0)W_{\scriptscriptstyle L}(y-x)\right],
\label{30}
\end{equation}
where we have used the fact that the local propagator $\Pi_{\scriptscriptstyle L}(x-y)$ corresponds to the time-ordered product between two fields $\phi(x)$ and $\phi(y)$. Because of the time-derivative component of the d'Alembertian in the exponential function $f(\Box_x)$, it is clear that the  propagator cannot maintain the same causal structure of the Feynman propagator of the standard local field theory. 

We now want to find the explicit form of the  propagator in the coordinate-space, and in order to do so we need to understand how to deal with the differential operators of infinite order. By using the identity~\footnote{The identity in Eq.\eqref{31} holds in flat spacetime as $[\partial_{x^0}^2,\nabla_{\vec{x}}^2]=0.$ In curved spacetime one has to deal with covariant derivatives and $\Box=g_{\mu \nu}\nabla^{\mu}\nabla^{\nu}$, so that the simple decomposition in Eq.\eqref{31} is not possible.}
\begin{equation}
\Box_x^n=(-\partial_{x^0}^2+\nabla_{\vec{x}}^2)^n=\sum\limits_{p=0}^{n}{n\choose p}\left(-\partial_{x^0}^2\right)^{(p)}\left(\nabla_{\vec{x}}^2\right)^{(n-p)}\,,\label{31}
\end{equation}
and the generalized Leibniz product-rule,
\begin{equation}
\partial_{x^0}^{(2p)}\left[g(x^0)h(x^0)\right]=\sum\limits_{q=0}^{2p}{2p\choose q}\partial_{x^0}^{(q)}g(x^0)\partial_{x^0}^{(2p-q)}h(x^0),\label{32}
\end{equation}
we can manipulate the expression in the last line of Eq.\eqref{30} and obtain:
\begin{equation}
\!\!\begin{array}{ll}
e^{f(\Box_x)}\left[\theta(x^0-y^0)W_{\scriptscriptstyle L}(x-y)\right]= \displaystyle \sum\limits_{n=0}^{\infty}\frac{f_n}{n!}\Box_x^n\left[\theta(x^0-y^0)W_{\scriptscriptstyle L}(x-y)\right] &\\
\begin{array}{ll}
&\,\,\,\,\,\,\,\,\,\,\,\,\,\,\,\,\,\,\,\,= \displaystyle \theta(x^0-y^0) W(x-y)+ \displaystyle \sum\limits_{q=1}^{\infty}\sum\limits_{n=0}^{\infty}\frac{f_n}{n!}\sum\limits_{p=0}^{n}{n\choose p} {2p\choose q}i^q \theta(2p-q) \partial_{x^0}^{(q-1)}\delta(x^0-y^0) \\
& \,\,\,\,\,\,\,\,\,\,\,\,\,\,\,\,\,\,\,\,\,\,\,\,\,\,\,\,\,\,\,\,\,\,\,\,\,\,\,\,\,\,\,\,\,\,\,\,\,\,\,\,\,\,\,\,\,\,\,\,\,\,\,\,\,\,\,\,\,\,\,\,\,\,\,\,\,\,\,\,\,\,\,\,\displaystyle \times \int \frac{d^3k}{(2\pi)^3}\frac{e^{ik\cdot (x-y)}}{2\omega_{\vec{k}}}(-\vec{k}^2)^{n-p}\omega_{\vec{k}}^{2p-q},
\end{array}
\end{array}\label{33}
\end{equation}
where in the last equality we have introduced the step-function $\theta(2p-q)$, so that we can extend the summation over $q$ up to infinity. We can now note that the identity
\begin{equation}
\displaystyle \frac{1}{q!}\frac{\partial^{(q)}e^{f(-k^2)}}{\partial k^{0(q)}}=  \displaystyle \sum\limits_{n=0}^{\infty}\frac{f_n}{n!}\sum\limits_{p=0}^{n}{n\choose p} {2p\choose q}\theta(2p-q)(k^0)^{2p-q}(-\vec{k}^2)^{n-k}.
\label{34}
\end{equation}
allows us to rewrite Eq.\eqref{33} as follows:
\begin{equation}
\!\!\begin{array}{ll}
\displaystyle e^{f(\Box_x)}\left[\theta(x^0-y^0)W_{\scriptscriptstyle L}(x-y)\right]=  \theta(x^0-y^0) W(x-y) & \\
\,\,\,\,\,\,\,\,\,\,\,\,\,\,\,\,\,\,\,\,\,\,\,\,\,\,\,\,\,\,\,\,\,\,\,+ \displaystyle i\sum\limits_{q=1}^{\infty}\frac{i^{q-1}}{q!}\partial_{x^0}^{(q-1)}\delta(x^0-y^0) \int \frac{d^3k}{(2\pi)^3}\frac{e^{ik\cdot (x-y)}}{2\omega_{\vec{k}}}\left.\frac{\partial^{(q)}e^{f(-k^2)}}{\partial k^{0(q)}}\right|_{k^0=\omega_{\vec{k}}}. &
\end{array}
\label{35}
\end{equation}
Following the same steps for the second term in Eq.\eqref{30}, one has
\begin{equation}
\!\!\begin{array}{ll}
\displaystyle e^{f(\Box_x)}\left[\theta(y^0-x^0)W_{\scriptscriptstyle L}(y-x)\right]= \theta(y^0-x^0) W(y-x) & \\
\,\,\,\,\,\,\,\,\,\,\,\,\,\,\,\,\,\,\,\,\,\,\,\,\,\,\,\,\,\,\,\,\,\,\,- \displaystyle i\sum\limits_{q=1}^{\infty}\frac{i^{q-1}}{q!}\partial_{x^0}^{(q-1)}\delta(x^0-y^0) \int \frac{d^3k}{(2\pi)^3}\frac{e^{ik\cdot (y-x)}}{2\omega_{\vec{k}}}\left.\frac{\partial^{(q)}e^{f(-k^2)}}{\partial k^{0(q)}}\right|_{k^0=\omega_{\vec{k}}}. &
\end{array}\label{36}
\end{equation}
We can now substitute Eqs.\eqref{35},\eqref{36} into Eq.\eqref{30}, and obtain a very interesting expression for the  propagator~\footnote{Eq.\eqref{37} is in agreement with the result obtained in Ref.\cite{tomboulis2015}, where the author has followed a different procedure.}:
\begin{equation}
\begin{array}{rl}
\Pi(x-y)= & \displaystyle \theta(x^0-y^0)W(x-y)+\theta(y^0-x^0)W(y-x) \\
& +\displaystyle i\sum\limits_{q=1}^{\infty}\frac{i^{q-1}}{q!}\partial_{x^0}^{(q-1)}\delta(x^0-y^0)[W^{(q)}(x-y)-W^{(q)}(y-x)], 
\end{array}
\label{37}
\end{equation}
where we have defined
\begin{equation}
\begin{array}{rl}
W^{(q)}(x-y):= & \displaystyle \int \frac{d^3k}{(2\pi)^3}\frac{e^{ik\cdot (x-y)}}{2\omega_{\vec{k}}}\left.\frac{\partial^{(q)}e^{f(-k^2)}}{\partial k^{0(q)}}\right|_{k^0=\omega_{\vec{k}}}\\
= & \displaystyle \int \frac{d^4k}{(2\pi)^3}e^{ik\cdot (x-y)}\theta(k^0)\delta(k^2+m^2)\frac{\partial^{(q)}e^{f(-k^2)}}{\partial k^{0(q)}}.
\end{array}
\label{38}
\end{equation}
From Eq.\eqref{37} it is clear that the  propagator is not just a time ordered product, but it also has an extra term that breaks the causal structure of the local Feynman propagator: this is a first example of causality violation induced by non-local interactions, as already been shown in Ref.~\cite{tomboulis2015}. In the standard local quantum field theory, the time-ordered product corresponds to the Feynman causal propagator that is constructed such that particles with positive-energy travels forward in time, while particles with negative energy (anti-particles) travel backwards in time. Such a structure is not preserved in infinite derivative field theory and causality is violated within $1/M_s$. For energies below $M_s$, the form factor reduces to $e^{-f(\Box)}\rightarrow 1$, and hence reaches the local field theory limit in the IR. In Section \ref{CV}, we will quantify the violation of causality in more detail. We can also define causal and non-causal (or acausal) parts of the propagator in Eq.\eqref{37}, as follows:
\begin{equation}
\Pi_{\scriptscriptstyle c}(x-y)=\theta(x^0-y^0)W(x-y)+\theta(y^0-x^0)W(y-x)
\label{39}
\end{equation}
and~\footnote{Note that in the non-causal term $\Pi_{\scriptscriptstyle nc}(x-y)$ there is an infinite number of contact terms that cannot be absorbed through counterterms, thus they will still be there once the theory is renormalized \cite{tomboulis2015}.}
\begin{equation}
\Pi_{\scriptscriptstyle nc}(x-y)=i\sum\limits_{q=1}^{\infty}\frac{i^{q-1}}{q!}\partial_{x^0}^{(q-1)}\delta(x^0-y^0)[W^{(q)}(x-y)-W^{(q)}(y-x)],
\label{40}
\end{equation}
so that the  propagator in Eq.\eqref{30} can be rewritten as~\footnote{The concept of propagator assumes physical meaning only when we consider propagation between two interaction vertices; thus, such a causality violation does not appear at the level of free-theory within infinite derivative theory, but only when the interaction is switched on.}
\begin{equation}
\Pi(x-y)=\Pi_{\scriptscriptstyle c}(x-y)+\Pi_{\scriptscriptstyle nc}(x-y).
\label{41}
\end{equation}
Since for free-fields $W(x-y)=W_{\scriptscriptstyle L}(x-y)$, one has $\Pi_{\scriptscriptstyle c}(x-y)=\Pi_{\scriptscriptstyle L}(x-y)$.



\section{Causality}\label{CV}

In this section we will explicitly show that the presence of non-local interactions violate causality in a region whose size is given by $l_s\simeq1/M_s$ in coordinate space, and for momenta $k^2> M_s^2$ in momentum space.

\subsection{A brief reminder}

Let us consider a real scalar field $\phi(x^0,\vec{x})$ that evolves by means of a differential operator $F(\Box)$ in presence of a source $j(x^0,x)$, so that it satisfies the following differential equation:
\begin{equation}
F(\Box)\phi(x^0,\vec{x})=-j(x^0,\vec{x}).
\label{44}
\end{equation}
A formal solution to Eq.~\eqref{44} is given by
\begin{equation}
\phi(x^0,\vec{x})=\phi_o(x^0,\vec{x})+i\int dy^0d^3y G(x^0-y^0,\vec{x}-\vec{y})j(y^0,\vec{y}),
\label{45}
\end{equation}
where $\phi_o(x^0,\vec{x})$ is the solution of the homogeneous equation, and $G(x^0-y^0,\vec{x}-\vec{y})$ is the Green function of the differential operator $F(\Box)$, defined by
\begin{equation}
F(\Box)G(x^0-y^0,\vec{x}-\vec{y})=i\delta(x^0-y^0)\delta^{(3)}(\vec{x}-\vec{y}).
\label{46}
\end{equation}
A system whose evolution is governed by Eq. \eqref{44} is said to be {\it causal} if the corresponding Green function $G(x^0,\vec{x})$ can be chosen, such that
\begin{equation}
G(x^0,\vec{x})=0\,,\,\,\,{\rm if} \,\,\,\,x^0<0.
\label{47}
\end{equation}
The statement in Eq.~\eqref{47} means that a physical system cannot respond to an interaction-source before the source was turned on~\footnote{Note that such a definition of causality in terms of the Green function, not only holds for classical fields, but also for the expectation value of quantum fields in presence of a source, $\left\langle \phi(x)\right\rangle_j$.}. The previous definition of causal response holds for both relativistic and non-relativistic systems. A stronger version of the condition in Eq.\eqref{47} is given by the concept of {\it sub-luminality} \cite{nicolis}, which is a property that has to be satisfied by any relativistic system. A physical system is said to be {\it sub-luminal}, if the Green function $G(x^0,\vec{x})$ is causal and also vanishes outside the light cone, i.e.
\begin{equation}
G(x^0,\vec{x})=0\,,\,\,\,{\rm if} \,\,\,\,x^0<|\vec{x}|.
\label{48}
\end{equation}
In this paper,  by {\it causality} we will also refer to the concept of sub-luminality, namely a causal system will be characterized by a vanishing Green function for space-like separations. Such a Green function is often indicated with a subscript "$R$" due to its retarded behavior, and we will use the symbol $G_{\scriptscriptstyle L,R}$ in the case of the standard local field theory. Another definition of causality is given through the commutator of two fields evaluated in two different space-time points. From a physical-measurement point of view, to preserve causality, we would require that the commutator of the two observables has to vanish outside the lightcone, i.e. for space-like separations. For a real scalar field, such a property can be formulated in the following way~\footnote{Let us remind that in the mostly positive metric signature $(x-y)^2>0$ stands for space-like separation and $(x-y)^2<0$ for time-like separation. In the mostly negative convention we would have had the opposite situation.}:
\begin{equation}
\left[\phi(x),\phi(y)\right]=0\,,\,\,\,{\rm if} \,\,\,\,(x-y)^2>0.
\label{49}
\end{equation}
When two observables commute, it means that they can be measured simultaneously, i.e. namely one measurement cannot influence the other. If the condition in Eq.\eqref{49} is violated, there would be correlations between the two measurements performed at two different spacetime points with space-like separation, implying transmission of information at a speed faster than light, thus violating causality. The property in Eq.\eqref{49} is called {\it local commutativity}, or sometime {\it microcausality}.

Note that the two conditions of causality given in terms of the Green function, see Eq.~(\ref{48}), and local commutativity, see Eq.~(\ref{49}), are closely related in local field theory. Let us consider a Hamiltonian interaction between a real scalar field $\phi(x)$ and a source $j(x)$, $H_{\rm int}=\int d^3xj\phi$. Consider an initial configuration with a vacuum state at a time $y^0=-\infty$ and then switch on the source at a later time. The expectation value of $\phi$ at a spacetime point $(x^0,\vec{x})$, with $x^0>y^0$, can be calculated in the interaction picture, and it is given by \cite{nicolis}
\begin{equation}
\begin{array}{rl}
\left\langle \phi(x)\right\rangle_j= & \displaystyle \langle 0|e^{i\int_{-\infty}^{x^{\scriptscriptstyle 0}} dy^0d^3y j(y)\phi(y)}\phi(x) e^{-i\int_{-\infty}^{x^{\scriptscriptstyle 0}} dy^0d^3y j(y)\phi(y)}|0\rangle\\
= & \displaystyle \langle \phi(x)\rangle_{j=0} -\int d^4yj(y)i\theta(x^0-y^0)\left\langle 0\left| \left[\phi(x),\phi(y)\right]\right| 0\right\rangle+\cdots
\end{array}
\label{50}
\end{equation}
where the dots stand for higher order contributions in the interaction-source term. By comparing Eq.\eqref{45} with Eq.\eqref{50} we can identify $\phi_o(x)=\langle \phi(x)\rangle_{j=0}$, and also note that in local field theory, the expectation value of the commutator between the two fields is related to the retarded Green function through the following relation:
\begin{equation}
G_{\scriptscriptstyle L,R}(x-y)=-\theta(x^0-y^0)\left\langle 0\left| \left[\phi(x),\phi(y)\right]\right| 0\right\rangle.
\label{51}
\end{equation}
Hence, if the commutator vanishes for space-like separations, the interaction-source can only generate non-zero modes inside its future lightcone, and thus the definition of causality given in terms of the Green function is consistent with the local commutativity condition. For completeness, we can also write the analog of Eq.\eqref{51} for the advanced Green function:
\begin{equation}
G_{\scriptscriptstyle L,A}(x-y)=\theta(y^0-x^0)\left\langle 0\left| \left[\phi(x),\phi(y)\right]\right| 0\right\rangle.
\label{51.2}
\end{equation}
%


\subsection{Acausal Green functions in infinite derivative field theory}\label{NLAGF}

We have already given an example of causality violation in Section \ref{IDA}, where we have shown that the propagator is not simply given by a time-ordered product, but it has an extra non-causal term which becomes relevant inside the non-local region. We now want to show that the presence of non-local interactions leads inevitably to a violation of causality inside the region  $\sim1/M_s$. In particular, we wish to show explicitly that the non-local analog of the retarded Green function, $e^{f(\Box)}[G_{\scriptscriptstyle L,R}(x^0,\vec{x})],$ is not vanishing outside the light-cone. We will simply indicate the non-local analog of the retarded Green function with the symbol $G_{\scriptscriptstyle R}$, meaning that it is a non-local quantity, while in presence of the subscript "$L$" we would refer to local quantities. 

Let us remind that in local quantum field theory the retarded Green function is defined in terms of its Fourier transform as
\begin{equation}
-iG_{\scriptscriptstyle L,R}(x-y)=\int\limits_{C_R}\frac{d^4k}{(2\pi)^4}\frac{e^{ik\cdot (x-y)}}{(k^0)^2-\vec{k}^2-m^2}, 
\label{65}
\end{equation}
where the integration contour $C_R$ is given by the real axis where both the poles: $\pm \omega_{\vec{k}}=\pm \sqrt{\vec{k}^2+m^2}$ are avoided from above with two semi-circles. By evaluating the integral in Eq.\eqref{65} in the massless case, one obtains the retarded Green function in coordinate space:
\begin{equation}
-iG_{\scriptscriptstyle L,R}(x-y)= \frac{1}{2\pi}\theta(t)\delta(\rho)=\frac{1}{4\pi r}\delta(t-r), 
\label{66}
\end{equation}
where $t=x^0-y^0$, $\vec{r}=\vec{x}-\vec{y}$ and $\rho=t^2-r^2$. From Eq.\eqref{66} it is obvious that the retarded propagator is vanishing outside the light-cone, i.e. in the region $t<r$.

We now want to treat the case of infinite derivative field theory and explicitly see that the retarded Green function shows an acausal behavior due to non-local interactions. By following the steps in Eqs.\eqref{35} and \eqref{36} together with Eq.\eqref{51}, one can write the non-local retarded Green function as follows 
\begin{equation}
G_R(x-y)= e^{f(\Box_{x})}G_{\scriptscriptstyle L,R}(x-y)
= -\theta(x^0-y^0)\left\langle 0\left|[\phi(x),\phi(y)]\right|0\right\rangle-\Pi_{nc}(x-y), 
\label{64}
\end{equation}
Note the presence of the acausal (non-causal) term $\Pi_{nc}$ introduced in Eq.\eqref{40}. In particular, we will consider  form-factors with polynomial exponents as in Eq.\eqref{eq:14}, and for this specific choice we will see which is the form of $\Pi_{nc}$.  

First of all, note that such  form factors are divergent at infinity along some directions in the complex plane $k^0$: for example, it can happen that they diverge at $-\infty$ and $+\infty$ along the real axis making it impossible to compute the integral in Eq.\eqref{66} in Minkowski signature. These kind of divergences make also impossible to define the usual Wick-rotation; this is one of the mathematical reason why in infinite derivative field theory one has to define all amplitudes in the Euclidean space, and in the end of the calculation go back to Minkowski signature by analytic continuation. Below we will give a more detailed discussion about this last observation.

However, in the case of the exponential choice in Eq.\eqref{eq:14} with even powers $n$ the non-local form-factor does not diverge along the real axis at infinity, and we can {\it still} compute the principal value of the integral in Minkowski signature. Therefore, in this subsection we will consider the following form factors~\footnote{Form factors with odd powers of $\Box$ can be computed in the region $\sim1/M_s$ once we go to the Euclidean signature, where one has a very interesting scenario in which all the Euclidean Green functions turn out to be non-singular at the Euclidean origin (lightcone surface in Minkowski signature) for any power $n$. See Section \ref{EU}, where we will consider the case for $n=1$. However, as we will emphasize in Section \ref{ME}, because of the presence of acausal effects inside the non-local region, all Green functions, with any power $n$, can be physically interpreted {\it only} in Euclidean signature for $|x-y|\leq1/M_s.$}:  
\begin{equation}
e^{-f(\Box)}=e^{\left(\frac{\Box}{M_s^{2}}\right)^{2n}}\label{eq:66}
\end{equation}
and we will work in the massless case for simplicity, $m=0$.  The aim is to compute the following integral:
\begin{equation}
-iG_{\scriptscriptstyle R}(x-y)=\int\limits_{C_R}\frac{d^4k}{(2\pi)^4}\frac{e^{-\left(\frac{k^2}{M_s^{2}}\right)^{2n}}e^{ik\cdot (x-y)}}{(k^0)^2-\vec{k}^2}.
\label{67}
\end{equation}
The integral in Eq.\eqref{67} can be split into its principal value plus the contribution coming from the two semi-circles that avoid the two poles from above:
\begin{equation}
-iG_{\scriptscriptstyle R}=I_{\scriptscriptstyle PV}+I_{\scriptscriptstyle 2C},
\label{68}
\end{equation}
where $I_{\scriptscriptstyle 2C}$ can be calculated by using the residue theorem, and one can easily show that is equal to
\begin{equation}
I_{\scriptscriptstyle 2C}= \displaystyle \frac{1}{8\pi r}\left[\delta(t-r)-\delta(t+r)\right]
= \displaystyle \frac{1}{4\pi}\varepsilon(\rho)\delta(\rho).
\label{69}
\end{equation}
As for the principal value, one has
\begin{equation}
I_{\scriptscriptstyle PV}=\frac{1}{16i\pi^3}\frac{1}{r} \int\limits_{-\infty}^{\infty} kdk \,\,{\rm P.V.} \!\int\limits_{-\infty}^{\infty}dk^{0}\frac{e^{-\left(\frac{-k_0^2+k^2}{M_s^{2}}\right)^{2n}}}{k_0^2-k^2}\left(e^{i(kr-k^0t)}-e^{-i(kr+k^0t)}\right), \label{70}
\end{equation}
where $k\equiv |\vec{k}|$ and $\omega_{\vec{k}}=k$, as we are working in the massless case. Note that all information about non-locality is contained in the principal value $I_{\scriptscriptstyle PV}$, while $I_{\scriptscriptstyle 2C}$ is just a local contribution as it is evaluated at the residues, i.e. on-shell.

After some manipulations, one can show that the principal value in Eq.\eqref{70} can be recast in the following form~\footnote{See Appendix \ref{app-princ} for all the details of the calculation.}:
\begin{equation}
I_{\scriptscriptstyle PV}=\frac{1}{\pi^3}\frac{\partial}{\partial\rho}\left\lbrace \varepsilon(\rho) \int\limits_{0}^{\infty}\frac{d\zeta}{\zeta}e^{-\frac{\zeta^{4n}}{M_s^{4n}\rho^{2n}}}\left[K_0(\zeta)+\frac{\pi}{2}Y_0(\zeta)\right]\right\rbrace,
\label{71.2}
\end{equation}
where $\varepsilon(\rho)$ is equal to $+1$ if $\rho>0$ (time-like separation), while it is $-1$ if $\rho<0$ (space-like separation); $Y_0$ and $K_0$ are Bessel functions of the second kind and the modified Bessel function, respectively.
\begin{figure}[t]
	\includegraphics[scale=0.50]{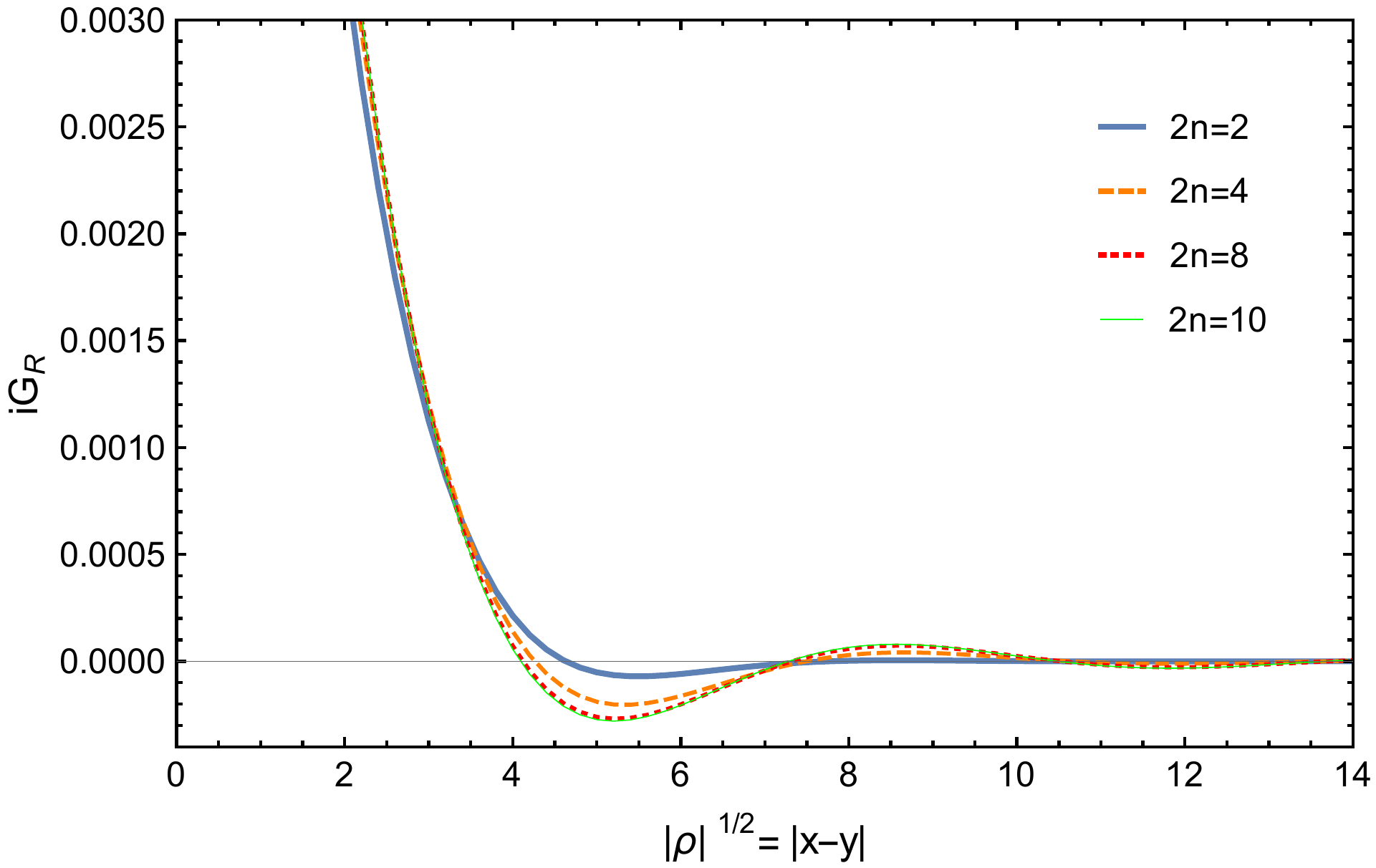}
	\centering
	\protect\caption{In this plot we have shown the behavior of the non-local analog of the retarded Green function as a function of the space-like distance $|\rho|^{1/2}=|x-y|$, $\rho<0,$ for several values of the power in the exponent: $2n=2$ (continuous thick blue line), $2n=4$ (dashed orange line), $2n=8$ (dotted red line) and $2n=10$ (continuous thin green line). The first two cases can be computed analytically and expressed in terms of the Meijer-G functions (see Eq.\eqref{77} for the case $2n=2$), while the last two cases have been obtained numerically. We have set $M_s=1$ as we are only interested in the qualitative behavior of the functions. It is evident that for small distances non-locality is relevant and we have an acausal behavior, but as soon as $|\rho|^{1/2}$ increases non-locality becomes less important and the Green function tends to a zero value recovering the local result, as expected. The oscillation-effects induced by non-locality increase with the power $2n$.}
\end{figure}

We can now find an explicit form for the integral in Eq.\eqref{71.2}, for example we can consider the power $2n=2$. In such a case the integral can be computed and expressed in terms of the Meijer-G functions \cite{luke}, so that the acausal Green function in Eq.\eqref{67},\eqref{68} reads
\begin{equation}
\begin{array}{ll}
-iG_{\scriptscriptstyle R}= \displaystyle \frac{1}{4\pi}\varepsilon(\rho)\delta(\rho) 
\displaystyle +\frac{1}{2\pi^4}\frac{\varepsilon(\rho)}{\rho}\left\lbrace G_{2,5}^{4,1}\left(\begin{array}{l}
0\\
0,0,\frac{1}{2},\frac{1}{2}
\end{array}\Biggr|\frac{M_{s}^{4}\rho^{2}}{256}\right)\!+\!2\pi^{2}G_{3,6}^{4,1}\left(\begin{array}{l}
0,-\frac{1}{4},\frac{1}{4}\\
0,0,\frac{1}{2},\frac{1}{2},-\frac{1}{4},\frac{1}{4}
\end{array}\Biggr|\frac{M_{s}^{4}\rho^{2}}{256}\right)\right\rbrace\!.
\end{array}
\label{77}
\end{equation}
From Eq.\eqref{77} it follows that the  Green function $G_{\scriptscriptstyle R}$ is not vanishing for space-like separation ($\rho<0$). In Fig. 1 we have plotted such a Green function for $\rho<0$ so that it is very clear that it assumes values different from zero, but for large value of $\rho$, i.e. for $M_s^{2}\rho\rightarrow\infty$, $G_{\scriptscriptstyle R}\rightarrow 0$, as expected. Thus, the violation of causality is restricted to the spacetime region of size approximatively given by $\sim1/M_s$. 

In the limit $M_s^2\rho\rightarrow\infty$ the integral in Eq. \eqref{71.2} reduces to:
\begin{equation}
\lim\limits_{\scriptscriptstyle M_s^2\rho\rightarrow\infty}I_{PV}=\frac{1}{8\pi r}[\delta(t+r)+\delta(t-r)],
\end{equation}	
so that the sum of the two contributions $I_{2C}+I_{PV}$ would recover the local result in Eq.\eqref{66}.

It is worth mentioning that for other values of $2n$ the integral in Eq.\eqref{71.2} also shows an acausal behavior, for example we have checked that in the case $2n=4$ the integral can be still expressed as a combination of Meijer-G functions; for larger values of $2n$ one can proceed numerically. In Fig. 1 we have also shown the behavior of the acausal Green function for $2n=4,8,10$. Moreover, the same procedure that we have used above can be used to compute the non-local analog of the advanced Green function, and it will lead to an opposite situation in which $G_{\scriptscriptstyle A}$ will be non-vanishing for time-like separations.


\subsection{Acausality for interacting fields}\label{acaus-appen}

We now wish to show that, due to the acausal feature of the Green functions, non-local interaction also implies the presence of acausality in the evolution of the fields; in particular we will see that the fields can depend acausally upon the initial data. Let us consider the Lagrangian for a real scalar field $\phi(x)$ with a  quartic interaction~\footnote{We could have considered any kind of interaction, but as an example we have chosen $\phi^4$. Moreover, we are working in the case in which the kinetic term is the standard Klein-Gordon operator and the interaction term is modified by the introduction of a  form factor; of course the results would be the same if we considered  non-local kinetic operator and local interaction vertices.} as an example:
\begin{equation}
\mathcal{L}=\frac{1}{2}\phi(x)(\Box-m^2)\phi(x) -\frac{\lambda}{4!}\left(e^{\frac{1}{2}f(\Box)}\phi(x)\right)^4,
\label{52}
\end{equation}
with corresponding field equations given by
\begin{equation}
(-\Box+m^2)\phi(x)=- \frac{\lambda}{3!}e^{\frac{1}{2}f(\Box)}\left(e^{\frac{1}{2}f(\Box)}\phi(x)\right)^3,
\label{53}
\end{equation}
where $\lambda$ is a dimensionless coupling constant. The field equation in Eq.\eqref{53} can be solved perturbatively by continuous iterations; the zeroth and first order are given by:
\begin{equation}
\begin{array}{rl}
(-\Box+m^2)\phi^{(0)}(x)= & 0,\\
(-\Box+m^2)\phi^{(1)}(x)= & -\displaystyle \frac{\lambda}{3!}e^{\frac{1}{2}f(\Box)}\left(e^{\frac{1}{2}f(\Box)}\phi^{(0)}(x)\right)^3,
\end{array}
\label{54}
\end{equation}
where the zeroth order is nothing but the homogeneous Klein-Gordon equation, whose local solutions are given by the free-field decomposition in Eq.\eqref{eq:16}, that we rewrite for clarity:
\begin{equation}
\phi^{(0)}(x^0,\vec{x})=\int \frac{d^3k}{(2\pi)^3}\frac{1}{\sqrt{2\omega_{\vec{k}}}}\left(a_{\vec{k}}e^{-i\omega_{\vec{k}}x^0 +i\vec{k}\cdot\vec{x}}+a^{*}_{\vec{k}}e^{i\omega_{\vec{k}}x^0-i\vec{k}\cdot\vec{x}}\right) \label{eq:55}.
\end{equation}
Note that the Fourier-transform of the field $\phi^{(0)}(x^0,\vec{x})$ with respect to the spatial coordinate $\vec{x}$, $\tilde{\phi}^{(0)}(x^0,\vec{k})$, can be expressed in terms of the initial field configuration, $\tilde{\phi}^{(0)}(0,\vec{k})$ and $\dot{\tilde{\phi}}^{(0)}(0,\vec{k})$, as follows~\footnote{Note that the symbol " $\cdot$ " means derivative with respect to $x^0.$}
\begin{equation}
\tilde{\phi}^{(0)}(x^0,\vec{k})=\tilde{\phi}^{(0)}(0,\vec{k}){\rm cos}(\omega_{\vec{k}}x^0)+\dot{\tilde{\phi}}^{(0)}(0,\vec{k})\frac{{\rm sin}(\omega_{\vec{k}}x^0)}{\omega_{\vec{k}}},    \label{eq:56}
\end{equation}
so that the free-field in Eq. \eqref{eq:55} can be rewritten as
\begin{equation}
\phi^{(0)}(x^0,\vec{x})=\int \frac{d^3k}{(2\pi)^3}e^{i\vec{k}\cdot \vec{x}}\left(\tilde{\phi}^{(0)}(0,\vec{k}){\rm cos}(\omega_{\vec{k}}x^0)+\dot{\tilde{\phi}}^{(0)}(0,\vec{k})\frac{{\rm sin}(\omega_{\vec{k}}x^0)}{\omega_{\vec{k}}}\right) \label{eq:57}.
\end{equation}

Let us now compute the variation of the free-field with respect to an initial field-configuration $\phi^{(0)}(y)$, with $y\equiv(0,\vec{y})$, such that the distance between $x$ and $y$ is space-like, $(x-y)^2>0$ (or, equivalently, $|\vec{x}-\vec{y}|>x^0$):
\begin{equation}
\displaystyle \frac{\delta\phi^{(0)}(x^0,\vec{x})}{\delta\phi^{(0)}(0,\vec{y})}= \displaystyle \int \frac{d^3k}{(2\pi)^3}e^{-ik\cdot(\vec{x}-\vec{y})}{\rm cos}(\omega_{\vec{k}}x^0)
= -\dot{\Delta}(x^0,\vec{x}-\vec{y}), 
\label{58}
\end{equation}
\begin{equation}
\displaystyle \frac{\delta\phi^{(0)}(x^0,\vec{x})}{\delta\dot{\phi}^{(0)}(0,\vec{y})}= \displaystyle \int \frac{d^3k}{(2\pi)^3}e^{-ik\cdot(\vec{x}-\vec{y})}\frac{{\rm sin}(\omega_{\vec{k}}x^0)}{\omega_{\vec{k}}}
=  -\Delta(x^0,\vec{x}-\vec{y}), 
\label{59}
\end{equation}
where we have used $\tilde{\phi}^{(0)}(0,\vec{k})=\int d^3x' e^{-i\vec{k}\cdot \vec{x}'}\phi^{(0)}(0,\vec{x}'),$ and $\delta\phi^{(0)}(0,\vec{x}')/\delta\phi^{(0)}(0,\vec{y})=\delta^{(3)}(\vec{x}'-\vec{y}).$ Note that we have obtained the Pauli-Jordan function introduced in the Subsection \ref{FRNLI} as a result of the functional differentiation. As we have already emphasized, $\Delta(x^0,\vec{x}-\vec{y})$ has only support inside the lightcone, and the same holds for its time-derivative; thus for space-like separations they vanish and causality is preserved at the level of free-theory.

Let us now consider the first order in perturbation, i.e. the differential equation in the second line of Eq.\eqref{eq:55}. The solution $\phi^{(1)}(x)$ is given by the sum of the homogeneous and the particular solutions, which we can indicate by $\phi^{(1)}_o(x)$ and $\phi^{(1)}_p(x)$, respectively:
\begin{equation}
\phi^{(1)}(x)=\phi^{(1)}_o(x)+\phi^{(1)}_p(x), \label{eq:60}
\end{equation}
where the particular solution has the physical information about the non-local interaction.
The homogeneous solution, $\phi^{(1)}_o(x),$ satisfies the same equation of $\phi^{(0)}(x)$, while the particular solution can be formally expressed in terms of the Green function as in Eq.\eqref{45}:
\begin{equation}
\phi^{(1)}_p(x^0,\vec{x})= \displaystyle \frac{i\lambda}{3!} \int dx'^0d^3x' e^{\frac{1}{2}f(\Box_{x'})}[G_{\scriptscriptstyle L,R}(x^0-x'^0,\vec{x}-\vec{x}')]\left(\phi^{(0)}(x'^0,\vec{x}')\right)^3,
\label{eq:61}
\end{equation}
where we have used the fact that $e^{\frac{1}{2}f(\Box)}\phi^{(0)}(x)=\phi^{(0)}(x)$,  at the zeroth order we have a free-field field propagation - satisfying the homogeneous Klein-Gordon equation, and we have also made use of the kernel representation of the exponential differential operator:
\begin{equation*}
e^{\frac{1}{2}f(\Box_x)}g(x)=\int d^4ye^{\frac{1}{2}f(\Box_y)}\delta^{(4)}(x-y)g(y).
\end{equation*}
All information about the presence of non-local interactions is contained in the particular solution; thus let us now calculate, as done for the zeroth order in Eqs.\eqref{58} and \eqref{59}, the variation of the field $\phi^{(1)}_p(x^0,\vec{x})$ with respect an initial field configuration $\phi^{(0)}(0,\vec{y})$, $\dot{\phi}^{(0)}(0,\vec{y})$\footnote{A similar computation was also done, for example, in Ref. \cite{eliezer} in the case of scalar field with cubic interaction, $[e^{\Box/M_s^2}\phi(x)]^3$, that can represent the interaction vertex for a dilaton field in string field theory.}:
\begin{equation}
\displaystyle \frac{\delta\phi^{(1)}_p(x^0,\vec{x})}{\delta\phi^{(0)}(0,\vec{y})}= \!\displaystyle -\frac{i\lambda}{2} \!\!\int \!dx'^0d^3x' e^{\frac{1}{2}f(\Box_{x'})}[G_{\scriptscriptstyle L,R}(x^0-x'^0,\vec{x}-\vec{x}')]\dot{\Delta}(x'^0,\vec{x}'-\vec{y})\left(\phi^{(0)}(x'^0,\vec{x}')\right)^2\!\!,
\label{62}
\end{equation}
\begin{equation}
\displaystyle \frac{\delta\phi^{(1)}_p(x^0,\vec{x})}{\delta\dot{\phi}^{(0)}(0,\vec{y})}=  \!\!\displaystyle -\frac{i\lambda}{2} \!\int \!dx'^0d^3x' e^{\frac{1}{2}f(\Box_{x'})}[G_{\scriptscriptstyle L,R}(x^0-x'^0,\vec{x}-\vec{x}')]\Delta(x'^0,\vec{x}'-\vec{y})\left(\phi^{(0)}(x'^0,\vec{x}')\right)^2\!\!.
\label{63}
\end{equation}
The action of the  differential operator on the local Green function in the integrals in Eqs.\eqref{62} and \eqref{63} makes the interacting field $\phi^{(1)}_p$ depending acausally upon the initial data: in fact, the integrals in Eq.\eqref{62} and \eqref{63} are not vanishing for space-like separations $|\vec{x}-\vec{y}|>x^0$ due to the non-zero contribution coming from the integration-region $x'^0<|\vec{x}'|$ as the function $e^{\frac{1}{2}f(\Box_{x'})}[G_{\scriptscriptstyle L,R}(x'^0,\vec{x}')]$ exhibits an acausal behavior, i.e. it is non-vanishing for the space-like separations $x'^0<|\vec{x}'|$, as shown in section 3 for the case $f(\Box)=(-\Box+m^2)^n/M_s^{2n}.$ Indeed, more explicitly one has the following situations.
\begin{itemize}
	
	\item In the local case, $f(\Box)=0$, the integrals in Eqs.\eqref{62},\eqref{63} get non-vanishing contributions coming from the integration region on which both the retarded Green function and the Pauli-Jordan function are non-zero. Such a region is defined by the following two inequalities:
	%
	$x^0-x'^0\geq|\vec{x}-\vec{x}'|,\,\,\,\,\,\,x'^0\geq|\vec{x}'-\vec{y}| $.
	%
	Moreover, since the initial time condition is $y^0=0$ and we are looking at the future evolution by means the non-local analog of the retarded Green function, the following inequality has to hold:
	%
	$x^0>x'^0>0$. 
	%
	We can now ask if the field $\phi^{(1)}_p(x)$ depends acausally upon the initial spacetime configuration $(0,\vec{y})$. One can easily show that by putting together the above inequalities, we obtain
	\begin{equation}
	x^0\geq|\vec{x}-\vec{y}|, \label{a.3}
	\end{equation}
	which implies that for space-like separation, $x^0<|\vec{x}-\vec{y}|,$ the integrals in  Eqs. \eqref{62},\eqref{63} are vanishing. Thus, in local field theory the field evolution turns out to be causal.
	
	\item In the case of non-local interactions, $f(\Box)\neq0$, the  Green function $G_R$ shows an acausal behavior, i.e. it is non-vanishing for space-like separations, thus we can not use the first inequality $x^0-x'^0\geq|\vec{x}-\vec{x}'|,\,\,\,\,\,\,x'^0\geq|\vec{x}'-\vec{y}| $, as done above for the local case. It follows that for space-like separation $x^0<|\vec{x}-\vec{y}|$ the functional derivatives in Eqs.\eqref{62},\eqref{63} do not vanish and the field can depend acausally on the initial data.
	
\end{itemize} 
The acausal behavior is confined to a region of size $\sim 1/M_s$, as it would be more explicit once a specific choice for the  form factor is made.



\subsection{Local commutativity violation}

We have shown that the presence of non-local interaction implies an acausal behavior of the Green functions, which in turn makes the interacting fields depending acausally upon the initial data. However, we have not investigated yet whether also the commutator between interacting fields is modified in such a way that the local commutativity condition is also violated. In Section \ref{WF}, we have shown that the commutator for free-fields is not modified by infinite derivatives, maintaining the same structure of the local theory. We now want to show that local commutativity is violated when non-local interaction is switched on.

Let us still consider the non-local $\phi^4$-theory as an example, i.e. the Lagrangian in Eq. \eqref{52}, and let us compute the commutator between two interacting fields by using the perturbative field solution $\phi(x)=\phi^{(0)}(x)+\phi^{(1)}(x)+\mathcal{O}(\lambda^2)$ introduced in the previous subsection. The commutator between two interacting fields up to order $\mathcal{O}(\lambda^2)$ is given by
\begin{equation}
\begin{array}{rl}
\displaystyle [\phi(x),\phi(y)]=&\displaystyle [\phi^{(0)}(x),\phi^{(0)}(y)]+[\phi^{(0)}(x),\phi^{(1)}(y)]\\[3mm]
& \displaystyle+[\phi^{(1)}(x),\phi^{(0)}(y)]+[\phi^{(1)}(x),\phi^{(1)}(y)]+\mathcal{O}(\lambda^3). \label{inter-commutator}
\end{array}
\end{equation}
Note that $[\phi^{(0)}(x),\phi^{(0)}(y)]$ obeys local commutativity as $\phi^{(0)}$ is not affected by non-locality, see Eqs.\eqref{54},\eqref{eq:24}. Moreover, from Eq.\eqref{eq:60} we know that $\phi^{(1)}=\phi^{(1)}_o+\phi^{(1)}_p,$ where $\phi^{(1)}_o$ is also not affected by non-locality being a homogeneous solution; thus all the information about the non-local modification of the commutator are taken into account by the terms involving $\phi^{(1)}_p.$ The second and third terms contributing to the commutator in Eq.\eqref{inter-commutator} are of the following form:
\begin{equation}
[\phi^{(0)}(x),\phi^{(1)}(y)]\sim \frac{i\lambda}{3!}\int d^4y'e^{\frac{1}{2}f(\Box_{y'})}[G_{\scriptscriptstyle L,R}(y-y')]\left(\phi^{(0)}(y')\right)^2[\phi^{(0)}(x),\phi^{(0)}(y')].\label{second-commut}
\end{equation}
\begin{itemize}
	\item In the local case, $f(\Box)=0,$ the integral in Eq.\eqref{second-commut} gets a non-vanishing contribution when the following two inequalities are satisfied:
	\begin{equation}
	y^0-y'^{0}\geq |\vec{y}-\vec{y}'|,\,\,\,\,\,\,y'^{0}\geq |\vec{y}'-\vec{x}| \label{comm-inequal-1}
	\end{equation}
	where we have taken $x^0=0$ without any loss of generality. We can now notice that, by choosing $y^0>x^0=0,$ the last two inequalities together imply 
	\begin{equation}
	y^0\geq |\vec{y}-\vec{x}|,
	\end{equation}
	which means that the commutator in Eq.\eqref{second-commut} gets non-vanishing contributions only for either time-like or null separations, in local field theory.
	
	\item In the case of non-local interaction, $f(\Box)\neq 0,$ the first inequality in Eq.\eqref{comm-inequal-1} can not be used as the Green function is acausal and gives a non-vanishing contribution also for space-like separations. As a result, the commutator in Eq.\eqref{second-commut} will be non-vanishing for space-like separations. 
\end{itemize}
The same scenario also holds for the fourth term in Eq.\eqref{inter-commutator}, whose expression is of the following form:
\begin{equation}
\begin{array}{rl}
[\phi^{(1)}(x),\phi^{(1)}(y)]\sim & \displaystyle \left(\frac{i\lambda}{3!}\right)^2\int d^4xd^4ye^{\frac{1}{2}f(\Box_{x'})}[G_{\scriptscriptstyle L,R}(x-x')]e^{\frac{1}{2}f(\Box_{y'})}[G_{\scriptscriptstyle L,R}(y-y')]\\
& \times \displaystyle  \left(\phi^{(0)}(y')\right)^2\left(\phi^{(0)}(x')\right)^2[\phi^{(0)}(x'),\phi^{(0)}(y')].\label{fourth-commut}
\end{array}
\end{equation}
\begin{itemize}
	\item In the local case, $f(\Box)=0,$ the integral in Eq.\eqref{fourth-commut} gets a non-vanishing contribution when the following three inequalities are satisfied:
	\begin{equation}
	x^{0}-x'^{0}\geq |\vec{x}-\vec{x}'|,\,\,\,\,\,\,y^0-y'^{0}\geq |\vec{y}-\vec{y}'|,\,\,\,\,\,\,x'^0-y'^{0}\geq |\vec{x}'-\vec{y}'|. \label{comm-inequal-2}
	\end{equation}
	For simplicity, we can take $y^0>x^0=0,$ without any loss of generality, and we can notice that all together the inequalities in Eq.\eqref{comm-inequal-2} imply
	\begin{equation}
	y^0\geq |\vec{x}-\vec{y}|,
	\end{equation}
	which means that the commutator in Eq.\eqref{fourth-commut} gets non-vanishing contributions only for either time-like or null separations, in local field theory.
	
	\item In the case of non-local interaction, $f(\Box)\neq 0,$ the first two inequalities in Eq.\eqref{comm-inequal-2} can not be used as the Green functions are acausal and give non-vanishing contributions also for space-like separations. As a result, the commutator in Eq.\eqref{fourth-commut} will be non-vanishing for space-like separation, meaning a violation of the local commutativity condition. 
\end{itemize}
Note that we have only considered the commutator up to quadratic order in the coupling constant, but it is clear that local commutativity will be also violated at higher order in perturbation theory. 

In order to quantify the degree of local commutativity violation we need to perform the computation by specifying an explicit form for the Green function and the Pauli-Jordan function to see how the integrals in Eqs.(\ref{second-commut},\ref{fourth-commut}) behaves for space-like separation; but it will be subject of future works.



\subsection{Region of non-locality}

We have seen that in a Lorentz-invariant quantum field theory, non-local interactions can yield causality violation.
In the case of only time - or space-dependence the non-local region is $t<1/M_s$ or $r<1/M_s$, respectively. However, in $(3+1)$-dimensions acausality is confined in a  spacetime region defined by the following inequalities:
\begin{equation}
-\frac{1}{M_s^2}<(x-y)^2<\frac{1}{M_s^2}. \label{3.1}
\end{equation}
By looking at the double inequalities in Eq.\eqref{3.1}, there is some ambiguities suggesting that causality violation extends on macroscopic scales in the direction of the lightcone surface, i.e. for large values of both $t$ and $r$. Indeed, by looking at the structure of the Green function, being Lorentz invariant it will only depend on $\rho=t^2-r^2,$ and will be non-zero also for $r,t\rightarrow\infty$, with $t\lesssim r$. However, we would expect acausal effects to emerge only in the region $r,t<1/M_s$ when studying the evolution of a field in terms of non-local Green functions. 

Let us consider a field $\phi(x)$ evolving in presence of an interaction-source $j(x)$, so that its dynamics will be governed by the non-local Green function $G_R(x-y)$ through the following integral equation:
\begin{equation}
\phi(x)=\phi_o(x)+i\int d^4yG_R(x-y)j(y), \label{3.4}
\end{equation}
where $\phi_o$ resolves the homogeneous field equation. The non-local form factor can be moved on the source under the integral sign, so that the integral in Eq.\eqref{3.4} can be written as
\begin{equation}
\int_{B(x^0,\vec{x})} d^4yG_{R,L}(x-y)e^{f(\Box_y)}j(y), \label{3.4.2}
\end{equation}
where now the integration region $B(x^0,\vec{x}):=\left\lbrace (y^0,\vec{y}):|\vec{x}-\vec{y}|\leq x^0-y^0\right\rbrace $ has support inside the lightcone as $G_{R,L}(x-y)=0$ for $|\vec{x}-\vec{y}|> x^0-y^0$, thus we expect that {\it no} acausal effects extends on macroscopic scales $(t,r>1/M_s)$ along the direction of the lightcone surface. However, this apparent macroscopic acausality which seems to extend along the direction of lightcone surface needs to be further investigated and it will be subject of future works. 

Note also that the presence of non-local interaction plays a crucial role when studying the {\it initial value problem} for the field evolution in Eq.\eqref{3.4}. Indeed, as extensively discussed in Ref.\cite{tomboulis2015}, acausal effects are such that the existence of solutions to the initial value problem can be established, but the uniqueness is lost, which is due to the fact that to obtain a solution on a time interval $[t_0,t_1]$ one has to specify not only initial data for past delays but also for future delays, and the latter would be of the order of the scale of non-locality, $t_1+1/M_s$. From a physical point of view, such an acausal time delay cannot be measured because every measurement process would average over time-scales longer than $1/M_s$. 


\section{Euclidean prescription} \label{ME}

From a physical point of view the presence of acausal effects means that there is no concept of Minkowski spacetime in the non-local region, non-locality is such that we can not define the usual concepts of space and time. We can not define clock and rulers to make any kind of measurements inside $1/M_s.$ For this reason, we believe that defining physical quantities in Minkowski signature in such a region would not make sense from a physical point of view, but the appropriate way to proceed would be to define Euclidean amplitudes and Euclidean correlators. Indeed, in Euclidean space we do not have any concept of real time, all Euclidean distances are space-like by definition.

Such a physical argument also has a mathematical counterpart. As we have already briefly mentioned in the previous subsections, in infinite derivative field theory the  form-factors introduce some ambiguities when performing calculations of integrals in momentum space. For example, the exponential form-factors with polynomial exponents introduced in Eq.\eqref{eq:14} can always appear in loop-integral and amplitudes in the form $e^{-(k^2/M_s)^n}$ where $n$ is a positive integer. For example, for the calculation of either propagator or any other Green functions, one has to deal with integrals of the following type:
\begin{equation}
I(x)=\int\limits_{-\infty}^{\infty}dk^0\frac{e^{-\left(\frac{-k_0^{2}+\vec{k}^2}{M_s^{2}}\right)^n}e^{-ik^0x^0+i\vec{k}\cdot\vec{x}}}{k_0^2-\omega_{\vec{k}}^2}.
\end{equation}
It is easy to understand that the presence of the  form factor gives divergent contributions along certain directions in the complex plane $k^0$; for instance, we can consider as examples $n=1$ and $n=2$.

\begin{itemize}
	
	\item In the case $n=1$ one has:
	\begin{equation}
	e^{\frac{k_0^{2}-\vec{k}^2}{M_s^{2}}}e^{-ik^0x^0+i\vec{k}\cdot\vec{x}} \sim e^{\frac{{\rm Re}^2(k^0)}{M_s^{2}}}e^{\frac{{\rm Im}(k^0)x^0}{M_s^{2}}}e^{-\frac{{\rm Im}^2(k^0)}{M_s^{2}}},
	\end{equation}
	which diverges at infinity along the directions belonging to the region $|{\rm Re}(k^0)|>|{\rm Im}(k^0)|,$ while it converges to zero along the directions such that $|{\rm Re}(k^0)|\leq|{\rm Im}(k^0)|.$
	
	\item In the case $n=2$, the relevant contribution at infinity is given by:
	\begin{equation}
	e^{-\frac{(-k_0^{2}+\vec{k}^2)^2}{M_s^{4}}}e^{-ik^0x^0+i\vec{k}\cdot\vec{x}} \sim e^{-\frac{{\rm Re}^4(k^0)}{M_s^{4}}}e^{-\frac{{\rm Im}^4(k^0)}{M_s^{4}}}e^{6\frac{{\rm Im}^2(k^0){\rm Re}^2(k^0)}{M_s^{4}}},
	\end{equation}
	that only diverges along the directions ${\rm Im}(k^0)=\pm {\rm Re}(k^0)$, while in the rest of the complex plane it approaches to zero at infinity.
	
\end{itemize}

Note that such divergences make it almost always impossible to calculate integrals in Minkowski signature, for example the usual Feynman contour prescription does not work anymore, because the contribution coming from the semi-circle in either the lower or the upper half of the complex plane receive an infinite contribution at infinity. It implies that the usual Wick-rotation cannot be defined. Furthermore, in Minkowski signature the optical theorem is not satisfied for  amplitudes and unitarity seems to be lost \cite{carone}; however one can show that by working in Euclidean space and then analytically continuing the external momenta to Minkowski, the theory turns out to be unitary \cite{sen-epsilon,carone,Chin:2018puw,Briscese:2018oyx,Pius:2018crk}. An important property of such exponential form-factors is that they always go to zero along the imaginary axis directions, ${\rm Im}(k^0)\rightarrow\pm \infty$, so that amplitudes in Euclidean signature are well-defined and can be legitimately computed. 


\subsection{Euclidean 2-point correlation function}\label{EU}

In local quantum field theory one has to deal with infinities which need to be regularized in order to give physical meaning to the theory. There are at least three kind of divergences that one can encounter:
\begin{enumerate}
	\item UV divergences $(k\rightarrow\infty)$;
	
	\item IR divergences $(k\rightarrow0)$;
	
	\item lightcone singularities $(|x-y|\rightarrow0)$.
\end{enumerate}
In principle, one can cure IR and UV divergences but, even after the renormalization procedure has been applied, the lightcone singularity, which corresponds to the singularity at the origin in Euclidean space, still remains uncured. In this section we wish to compute the $2$-point correlation function in infinite derivative quantum field theory; as an example we will still consider $\phi^4$-theory. In particular, we want to analyze its behavior on the light-cone surface, or in other words at the Euclidean origin, and see whether non-local interactions can regularize the divergence at $(x-y)\rightarrow0$ from which the local theory suffers. For simplicity, we will focus on the form-factor $e^{f(\Box)}=e^{-(\Box-m^2)/M_s^{2}}$. As we have strongly stressed in Subsection \ref{ME}, we will be formulating our theory in the Euclidean space; thus let us consider the following Euclidean generating functional:
\begin{equation}
\mathcal{Z}[J]=\int\mathcal{D}\phi e^{-S_{\scriptscriptstyle E}[\phi]+\int d^4x J\phi},\label{eq:5.1}
\end{equation}
where $J(x)$ is the source-term and the Euclidean action is given by:
\begin{equation}
S_{\scriptscriptstyle E}[\phi]=\int d^4x\left(-\frac{1}{2}\phi(x)e^{-(\Box-m^2)/M_s^{2}}(\Box-m^2)\phi(x)+\frac{\lambda}{4!}\phi^4(x)\right).\label{5.2}
\end{equation}
The functional in Eq.\eqref{5.2} can be rewritten in the following way:
\begin{equation}
\mathcal{Z}[J]=\displaystyle e^{-\frac{\lambda}{4!}\int d^4x \left[\frac{\delta}{\delta J(x)}\right]^4}\mathcal{Z}_0[J]=\displaystyle e^{-\frac{\lambda}{4!}\int d^4x \left[\frac{\delta}{\delta J(x)}\right]^4}e^{\frac{1}{2}\int d^4xd^4y J(x)\Pi(x-y)J(y)},\label{5.3}
\end{equation}
where $\mathcal{Z}_0[J]$ is the free generating functional and $\Pi(x-y)$ is the  propagator in the Euclidean space where, now, $k\equiv(k_4,\vec{k})$ stands for the Euclidean momentum, with $k^4=-ik^0$, and $x\equiv(x^4,\vec{x})$ for the Euclidean coordinate, with $x^4=ix^0.$ 
We are interested in computing the $2$-point correlation function that is defined as
\begin{equation}
\mathcal{G}(x-y):=\left.\frac{\delta^2\mathcal{Z}[J]}{\delta J(x)\delta J(y)}\right|_{J=0}.\label{5.6}
\end{equation}
By expanding the exponential in Eq.\eqref{5.3}, we can compute perturbatively the correlator $\mathcal{G}(x-y)$; for instance up to the first order in $\lambda$, we obtain:
\begin{equation}
\mathcal{G}(x-y)=\Pi(x-y)-\frac{\lambda}{2}\Pi(0)\int d^4z\Pi(x-z)\Pi(z-y)+\mathcal{O}(\lambda^2),\label{5.7}
\end{equation}
where at zeroth order we have the free-propagator, while at the first order a tadpole contribution.

Let us start by analyzing the zeroth order of the perturbative expansion in Eq.\eqref{5.7}, i.e. the Euclidean propagator, in both massless and massive case\footnote{Both these examples comprise the case of odd power of $\Box$ that could not be computed in Minkowski signature, as discussed in section \ref{NLAGF}, and discussion surrounding Eq.\eqref{eq:66}.}.

\begin{itemize}
	
	\item In the massless case the Euclidean $2$-point function at zeroth order reads:
	\begin{equation}
	\left.\Pi_{\scriptscriptstyle}(x-y)\right|_{m=0}= \displaystyle \int \frac{d^4k}{(2\pi)^4}\frac{e^{-k^2/M_s^{2}e^{ik\cdot (x-y)}}}{k^2}=\displaystyle \frac{1}{4\pi^2(x-y)^2}\left(1-e^{-\frac{M_s^{2}(x-y)^2}{4}}\right).
	\label{5.8}
	\end{equation}
	First of all, note that in the limit $M_s^2(x-y)^2\rightarrow\infty$, we recover the local massless propagator,
	$\left.\Pi_{\scriptscriptstyle L}(x-y)\right|_{m=0}=\frac{1}{4\pi^2(x-y)^2}.$
	More importantly, in the limit in which non-locality becomes relevant, i.e. $M_s^{2}(x-y)^2\rightarrow0$, unlike in the local case the massless propagator in Eq.\eqref{5.8} does not diverge but it tends to a finite constant value:
	\begin{equation}
	\left.\Pi(0)\right|_{m=0}=\frac{M_s^2}{16\pi^2}.\label{5.11}
	\end{equation}
	The result in Eq.\eqref{5.11} is extremely important for what concerns the UV behavior of the theory. The quantity $\left.\Pi(0)\right|_{m=0}$ appears as a coefficient of the perturbative series in Eq.\eqref{5.7}, and in local field theory the renormalization problem arises because of the presence of divergent coefficients. Thus, we have seen a first concrete example of how non-local  interaction can improve the UV behavior of the theory. In particular, $\phi^4$-theory with non-local interaction becomes finite as discussed in Ref.\cite{Biswas:2014yia,Talaganis:2014ida}.
	
	\item As for the massive propagator, by using again the Schwinger parametrization for $1/(k^2+m^2)$, we can write
	\begin{equation}
	\Pi(x-y)= \displaystyle \int \frac{d^4k}{(2\pi)^4}\frac{e^{-(k^2+m^2)/M_s^{2}}e^{ik\cdot (x-y)}}{k^2+m^2}= \displaystyle \frac{m^{2}}{16\pi^2}\int\limits_{\frac{m^2}{M^2_s}}^{\infty}ds\frac{e^{-s}e^{-\frac{m^2(x-y)^2}{4s}}}{s^2}.
	\label{5.8.2}
	\end{equation}
	Although the integral in Eq.\eqref{5.8.2} cannot be solved in terms of elementary functions as in the massless case (Eq.\eqref{5.8}), it can be expressed in terms of the so called {\it cylindrical incomplete function} of Sonine-Schlaefli:
	\begin{equation}
	\Pi(x-y)= -\frac{1}{4\pi}\frac{m}{|x-y|}S_1(-\frac{m^2}{M_s^2},-\infty;im|x-y|),
	\label{5.8.3}
	\end{equation}
	where the Sonine-Schlaefli function is defined as \cite{cylindric} $$S_\nu(-p,-q;iz):=\frac{e^{-\frac{i\pi\nu}{2}}}{2\pi i}\left(\frac{z}{2}\right)^{\nu}\int\limits_p^q dt\,t^{-\nu -1}e^{-t-\frac{z^2}{4t}}.$$
	
	We can study the limit $(x-y)\rightarrow0$, and note that the massive propagator is non-singular at the Euclidean origin. Indeed, in this limit the integral in Eq.\eqref{5.8.2} gives
	\begin{equation}
	\Pi(0)=  \displaystyle \frac{M_s^2e^{-\frac{m^2}{M_s^2}}}{16\pi^2}\left[1+e^{\frac{m^2}{M_s^2}}{\rm Ei}\left(-\frac{m^2}{M_s^{2}}\right)\right],
	\label{5.12}
	\end{equation}
	where $${\rm Ei}(x):=-\int_{-x}^{\infty}dt\frac{e^{-t}}{t}$$ is the so called {\it exponential-integral function}. 
\end{itemize}

\subsubsection{First-order correction $\mathcal{O}(\lambda)$} \label{FO}

So far we have learnt that at the zeroth order in the perturbative expansion in Eq.\eqref{5.7} the $2$-point correlation function is regular at the Euclidean origin unlike in the local case where singularities are present. We now want to study the first order correction (tadpole) in Eq.\eqref{5.7} and see whether such a  regularization property is maintained. 
\begin{itemize}
	\item In the massive case, one can check numerically that the first order correction is non-singular at the Euclidean origin.
	
	\item In the massless case, the first order correction to the $2$-point Euclidean correlator is singular at the origin, as we will now show with an explicit calculation. However, this can be made non-singular by dressing the propagator; see below.
\end{itemize}
At the first order in perturbation theory, the $2$-point function for the massless case is given by
\begin{equation}
\begin{array}{rl}
\left.\mathcal{G}^{(1)}(x-y)\right|_{m=0}= & \displaystyle -\frac{\lambda}{2}\left.\Pi(0)\right|_{m=0}\int d^4z\left.\Pi(x-z)\right|_{m=0}\left.\Pi(z-y)\right|_{m=0}\\
= & -\displaystyle \frac{\lambda}{2}\left.\Pi(0)\right|_{m=0}\frac{1}{4\pi^2|x-y|}\int\limits_0^{\infty}dk\frac{e^{-\frac{2k^2}{M_s^{2}}}J_1(k|x-y|)}{k^2},
\end{array}\label{5.17}
\end{equation}
where we have moved to polar coordinates in 4-dimensions: $d^4k=k^3{\rm sin}\theta{\rm sin}^{2}\alpha dkd\theta d\alpha d\varphi$.
First of all, note that the integral in Eq.\eqref{5.17} has an IR divergence, as we can see more explicitly by introducing an IR cut-off $L$:
\begin{equation}
\begin{array}{ll}
\displaystyle \mathcal{G}^{(1)}(x-y)|_{m=0}= \displaystyle -\frac{\lambda}{2}\Pi(0)|_{m=0}\left\lbrace -\frac{1}{2\pi^2}\frac{1}{M_s^{2}(x-y)^2}\left(1-e^{-\frac{M_s^2(x-y)^2}{8}}\right)\right. & \\
\,\,\,\, \displaystyle \left. +\frac{1}{16\pi^2}{\rm Ei}\left(-\frac{M_s^{2}(x-y)^2}{8}\right) + \lim\limits_{L\rightarrow\infty}\left[ \frac{1}{4\pi^2}\frac{L}{|x-y|}\left(1-e^{-\frac{|x-y|}{4L}}\right)-\frac{1}{16\pi^2}{\rm Ei}\left(-\frac{|x-y|}{4L}\right)\right]\right\rbrace. &
\end{array}\label{5.18}
\end{equation}
It is obvious that in the limit $L\rightarrow\infty$, the IR divergence comes from
\begin{equation}
\lim\limits_{L\rightarrow\infty}{\rm Ei}\left(-\frac{|x-y|}{4L}\right)=\infty.\label{5.19}
\end{equation}
Note that the IR divergence comes together with the singularity at the Euclidean origin (or lightcone singularity in Minkowski signature) in the massless case - the correlator at the first order in $\lambda$ in Eq.\eqref{5.18} also diverges for $|x-y|\rightarrow0$.\footnote{The singularity at the Euclidean origin will also appear for any power of the d'Alembertian $\Box^n$, with any $n$; indeed, it is purely related to the infrared divergence that one has in the massless case. It so happens that infrared divergence and singularity at the Euclidean origin (or lightcone singularity) are mixed.}

We have seen that the first order correction seems to suggest that non-locality is not sufficient to regularize the singularity of the $2$-point correlation function in the massless case. However, as we will show below, if we consider the full correlator, namely taking into account all the quantum perturbative corrections through the so called dressed propagator, then we will see that the physical $2$-point function becomes regular at the origin.

\subsubsection{Dressed $2$-point correlation function}\label{DP}

Let us consider the Fourier transform $\mathcal{G}(k)$ of the correlator in Eq.\eqref{5.7}, in the more general case of massive scalar field and then we will also specialize to the massless case. It is well known that once one takes into account all the perturbative corrections to the $2$-point correlation function in momentum space one obtains the so called dressed-propagator that can be expressed in terms of the self-energy $\Sigma(k)$. Thus, we have \cite{Biswas:2014yia}\footnote{Note that the pole structure of the dressed propagator is described by the equation $k^2+m^2+\Sigma(k)e^{-\frac{k^2+m^2}{M_s^{2}}}=0$; see Section \ref{dress-unitar} for discussions.} 
\begin{equation}
\mathcal{G}(k)
= \displaystyle \Pi(k)\sum\limits_{n=0}^{\infty}(-1)^n\left[\Sigma(k) \Pi(k)\right]^n
=  \displaystyle \frac{\Pi(k)}{1+\Sigma(k)\Pi(k)}
=  \displaystyle \frac{e^{-\frac{k^2+m^2}{M_s^{2}}}}{k^2+m^2+\Sigma(k)e^{-\frac{k^2+m^2}{M_s^{2}}}}.
\label{5.21}
\end{equation}

\begin{figure}[t]
	\includegraphics[scale=0.525]{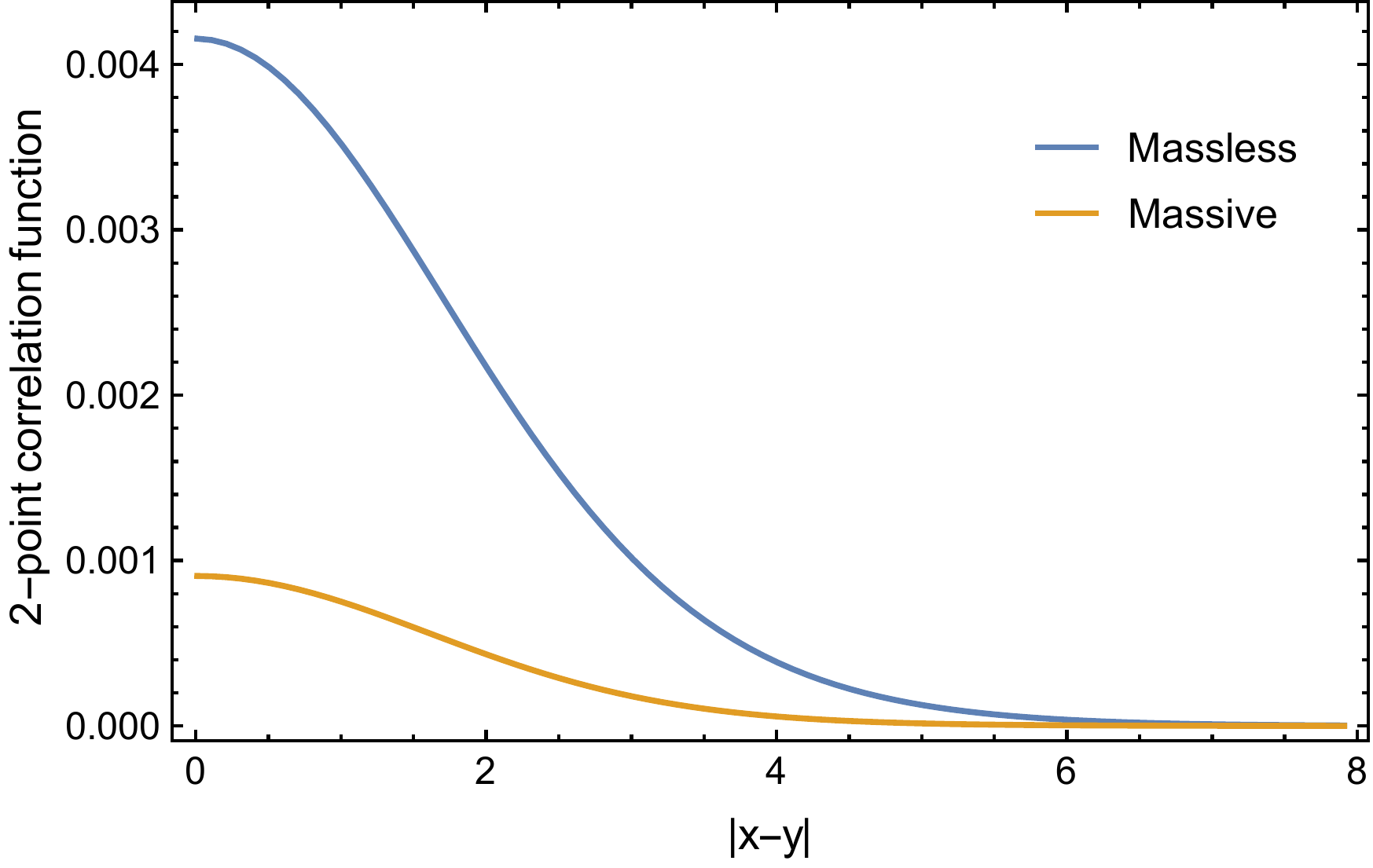}
	\centering
	\protect\caption{In this plot we have shown the behavior of the dressed $2$-point correlation function obtained by solving numerically the integral in Eq.\eqref{5.23}. The blue-line represent the massless case, while the orange line represents the massive case with $m=1$. We have set $M_s=1$ for simplicity, as we are only interested in the qualitative behavior around the origin. We can notice that for $|x-y|\rightarrow0$ the correlators tend to a finite value that of course differ for massless and massive cases.}\label{fig:3}
\end{figure}
The self-energy at $1$-loop is independent on the external momenta (tad-pole), and reads:
\begin{equation}
\Sigma=\lambda\int \frac{d^4p}{(2\pi)^4}\frac{e^{-\frac{(p^2+m^2)}{M_s^{2}}}}{p^2+m^2},\label{5.22}
\end{equation}
The integral in Eq.\eqref{5.22} turns out to be finite in both massless and massive cases; in fact it has the same expression of the Euclidean propagator evaluated at the origin in 
Eqs.\eqref{5.11} and \eqref{5.12} for massless and massive cases, respectively.

We are interested in the coordinate-space dressed-correlator, so we need to consider the following Fourier transform:
\begin{equation}
\mathcal{G}(x-y)= \displaystyle \int \frac{d^4k}{(2\pi)^4} \frac{e^{-\frac{k^2+m^2}{M_s^{2}}}e^{ik\cdot(x-y)}}{k^2+m^2+\Sigma(k)e^{-\frac{k^2+m^2}{M_s^{2}}}}
=  \displaystyle \frac{1}{4\pi^2|x-y|}\int\limits_{0}^{\infty}\frac{k^2J_1(k|x-y|)dk}{e^{\frac{k^2+m^2}{M_s^{2}}}(k^2+m^2)+\Sigma(k)}.
\label{5.23}
\end{equation}
The integral in Eq.\eqref{5.23} cannot be solved analytically, but we can calculate it numerically, and we note that the full $2$-point correlation function is non-singular at the Euclidean origin for both massive and massless cases. In Fig.\ref{fig:3}, we have shown the numerical solutions of the integral in Eq.\eqref{5.23} for both massive and massless cases. 


\section{Scattering amplitudes}\label{scattering-amplitudes}

In this section we will clarify that the amplitudes in infinite derivative quantum field theory can be well-defined and physically meaningful. First of all, note that some {\it apparent} ambiguities may appear when working with  amplitudes with the exponentials of the kind $e^{-(k^{2}/M_s^{2})^n}:$
\begin{enumerate}
	\item  $e^{-(k^{2}/M_s^{2})^{2n}}$: tree-level amplitudes with {\it even} power of the exponent would be always exponentially suppressed for both time-like and space-like momentum exchange;
	
	\item $e^{-(k^{2}/M_s^{2})^{2n+1}}$: tree-level amplitudes with {\it odd} power of the exponent would be exponentially suppressed {\it only} for space-like momentum exchange, but they blow up for time-like exchange, i.e. for $k^2<0.$
	
\end{enumerate}  
As a consequence, for both even and odd powers the {\it tree-level} scattering amplitudes turn out to be exponentially suppressed in the case of $t$- and $u$-channels, while the $s$-channel amplitude is exponentially suppressed only for even powers, but it blows up for odd powers as in this case the momentum exchange is time-like, giving a positive exponent in the exponential which causes the divergence for high energies. Such a divergent behavior appears when the amplitude is made of an internal propagator that connects two cubic vertices (e.g. $\phi^3$-theory).

However, this apparent unstable behavior appears for values of the energies beyond the cut-off $M_s,$ where we now know that no physical measurement can be made; moreover, such divergences only manifest at the level of the {\it bare propagator}. In fact, once all quantum corrections are taken into account through the {\it dressed propagator}, all the scattering amplitudes become exponentially suppressed, thus physically well-defined; see also for discussions in Ref.~\cite{talaganis-scatt}, where the tree-level scattering amplitudes were computed, and also the dressed vertices and the dressed propagator. The importance of using the dressed propagator, instead of the bare one, also arises when studying the renormalizability of non-local quantum field theories with infinitely many derivatives. As shown in Ref.\cite{Talaganis:2014ida}, the procedure of dressing the propagators ameliorates the UV aspects of the theory, making all loop-integrals finite.

Note that in the region of non-locality $\leq 1/M_s,$ or in momentum space, for momenta $k^2\geq M_s^2,$ we cannot define any classical concept of spacetime point, but vertices are smeared out such that the external legs and internal lines do {\it not} join in one point but they overlap in a region of size $1/M_s$. The crucial role is played by the {\it acausal} term $\Pi_{nc}$ defined in Eq.\eqref{40}, which implies causality violation in the vertices. In momentum space $\Pi_{nc}$ reads \cite{tomboulis2015}:
\begin{equation}
\begin{array}{ll}
\Pi_{nc}(k)=  \displaystyle \int d^4x\Pi_{nc}(x)e^{-ik\cdot x}&\\
\,\,\,\,\,\,\,=  \displaystyle i\sum\limits_{q=1}^{\infty}\frac{1}{q!}\frac{1}{2\omega_{\vec{k}}}\left[\left.\frac{\partial^{(q)}e^{f(-k^2)}}{\partial k^{0(q)}}\right|_{k^0=\omega_{\vec{k}}}(k^0-\omega_{\vec{k}})^{q-1}-\left.\frac{\partial^{(q)}e^{f(-k^2)}}{\partial k^{0(q)}}\right|_{k^0=-\omega_{\vec{k}}}(k^0+\omega_{\vec{k}})^{q-1}\right],&
\end{array}\label{vertex1}
\end{equation}
\begin{figure}[t]
	\includegraphics[scale=0.26]{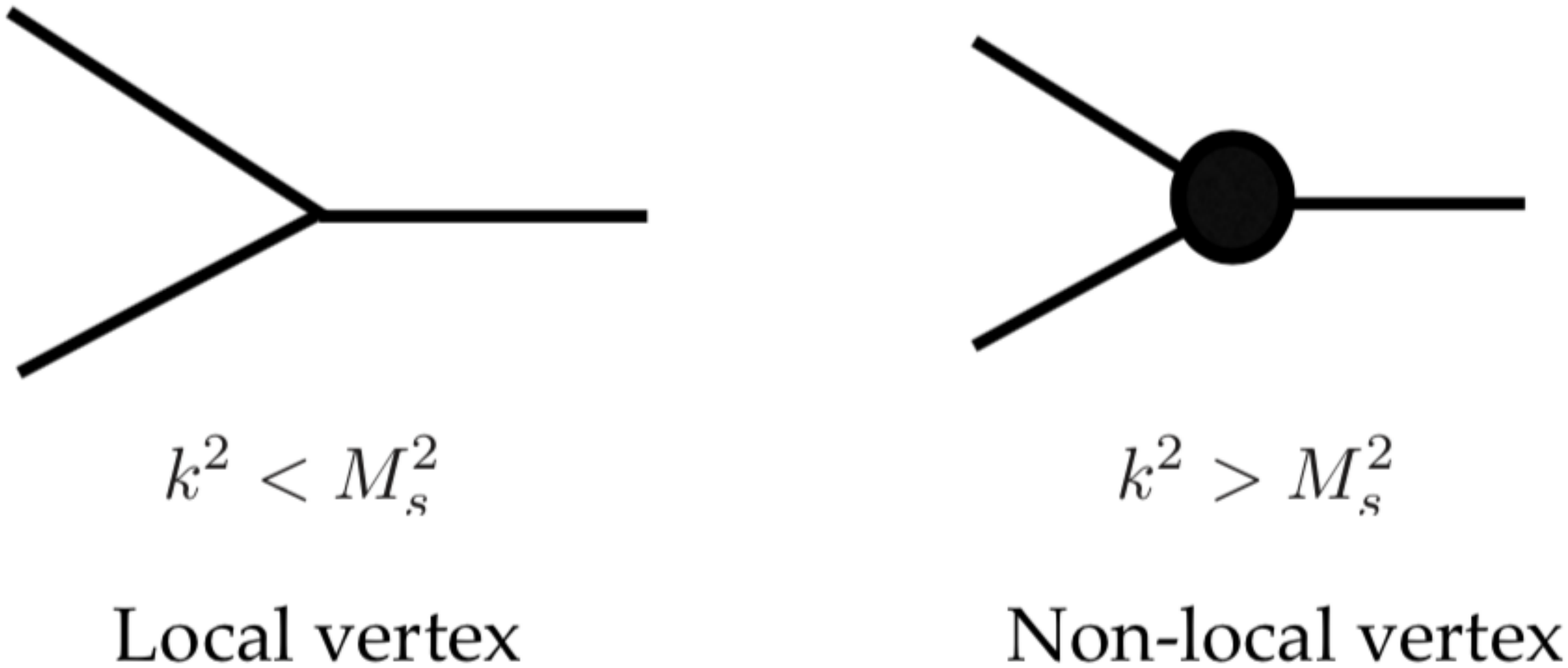}
	\centering
	\protect\caption{We have shown a pictorial illustration for local (left side) and non-local (right side) vertices. We can notice that above the scale of non-locality, $k^2\geq M_s^2,$ non-local interactions are such that the vertices are smeared out on a region of size $1/M_s$.}\label{fig:4}
\end{figure}
and it is evident that it has no poles and has no absorptive components, namely it is not made of on-shell intermediate states, but it is purely {\it off-shell}; indeed, it can be seen as a non-local vertex: $\Pi_{nc}(k)=iV(k).$ Thus, in momentum space the non-local propagator in Eq.\eqref{41} reads:
\begin{equation}
\Pi(k)=\Pi_c(k)+iV(k),\label{vertex2}
\end{equation}
where $\Pi_c(k)$ is causal and $V(k)$ tends to zero for $k^2\ll M_s^2$, but would be relevant for  $k^2\geq M_s^2$. Note that the non-local acausal part of the propagator, $iV(k)$, is made of infinite derivatives and this is the main cause of the smearing of the vertices, which are not point-like anymore\footnote{It is worth emphasizing that infinitely many derivatives can smear out point like source. In fact, by acting with infinite derivatives on a delta Dirac distribution, which has a point-like support, we obtain a non-point support \cite{Buoninfante:2018rlq}. For example, in the case of an exponential we obtain a Gaussian smearing:
	\begin{equation}
	e^{\alpha \partial_x^2}\delta(x)=\frac{1}{\sqrt{2\alpha}}e^{-\frac{x^2}{4\alpha}},
	\end{equation}
	where we have used the Fourier transform of the Dirac delta distribution.}. Thus, from Eqs.(\ref{vertex1},\ref{vertex2}) it is now more clear that in infinite derivative field theories non-locality and acausality manifest as {\it off-shell} effects, so that for momenta $k^2\geq M_s^2$ one has to consider any amplitudes as quantum and consistently take into account all perturbative quantum corrections.

In the standard local quantum field theory all internal lines of a Feynman diagram are seen as off-shell, while in infinite derivative quantum field theory the degree of {\it "off-shellness"} increases as also the vertices become non-local. In particular, there is no energy and momentum conservation in one single point,  as legs and internal lines overlap on a smeared region of size $1/M_s$, or in terms of momentum, $M_s.$ See Fig. \ref{fig:4} for an illustration of local and non-local vertices.

In this respect, bare amplitudes, which can be seen as classical amplitudes, do not make sense within the non-local regime, where we cannot define any classical concept of spacetime point, but instead vertices are smeared out. Indeed, what makes sense is the quantum scattering, and therefore the correct procedure will be always to consider the dressed vertices and dressed propagators irrespective of the cases of even powers $e^{-(k^{2}/M_s^{2})^{2n}},$ or odd powers $e^{-(k^{2}/M_s^{2})^{2n+1}}$, for any kind of amplitudes. We will compute these amplitudes for $\phi^3$ interaction.

\subsection{$s$- and $t$-channels}

In this subsection we will show that once all perturbative corrections are consistently taken into account for any channels and any power $n$ of the d'Alembertian, all scattering amplitudes are well-defined and have the same asymptotic behavior in the UV regime. As we have already mentioned above, some ambiguities can arise when considering odd powers, $e^{-(k^2/M_s^2)^{2n+1}},$ in the case of $s$-channel, where the momentum exchange is time-like, $k^2<0,$ giving a divergence at high energies, $k^2\gg M_s^2.$ We now want to explicitly show that by correctly dressing the propagator {\it no} such ambiguity would arise. For simplicity, we will consider the case of $\lambda \phi^3$-theory with a non-local kinetic operator, and work in the massless case. A generic tree-level scattering amplitude will be given by:
\begin{equation}
\mathcal{M}_n\sim \lambda^2\frac{e^{-(k^2/M_s^2)^{n}}}{k^2}.
\end{equation}
Once we dress the propagator, we will obtain (see Eq.\eqref{5.21})
\begin{equation}
\mathcal{M}_{n}\sim \lambda^2\frac{e^{-(k^2/M_s^2)^{n}}}{k^2+\Sigma_n(k) e^{-(k^2/M_s^2)^{n}}},\label{dress-amp}
\end{equation}
where the self-energy $\Sigma_n(k)$ for $\lambda \phi^3$-theory, for example at $1$-loop, reads
\begin{equation}
\Sigma_{n}(k)=\lambda^2\int \frac{dp^4}{(2\pi)^4} \frac{e^{-(p^2/M_s^2)^n}e^{-((p-k)^2/M_s^2)^n}}{p^2(p-k)^2}, \label{self-n}
\end{equation}
which for even powers of $n$ turns out to be always exponentially suppressed for any value of $k^2$, while for odd power of $n$ can blow up for time-like momenta, $k^2<0$. Note that, generally, integrals of the type in Eq.\eqref{self-n} can blow up for some values of the integration variable $p$; however, by working in Euclidean signature, $p^0=ip^4$ and $k^0=ik^4,$ the integrals can be computed, and after the computation is made the momentum $k$ can be analytically continued back to Minkowski signature, $k^4=-i k^0$; see for example for discussions Refs.\cite{sen-epsilon,witten,Biswas:2014yia}. Below we will make an explicit example for the case $n=1.$ 

Let us analyze $t$-channel and $s$-channel, for both even and odd powers. 

\subsubsection{Even powers $\Box^{2n} $}

Note that in the case of even powers, $e^{-(k^2/M_s^2)^{2n}}$, the high energy behavior of the scattering amplitudes is the same for both bare and dressed propagator. Indeed, in the UV regime, $k^2/M_s\rightarrow \infty$ we have the following asymptotic behavior for the dressed amplitude in Eq.\eqref{dress-amp} with even powers:
\begin{equation}
\mathcal{M}_{2n}\sim e^{-(k^2/M_s^2)^{2n}},
\end{equation}
as $\Sigma_{2n}(k)e^{-(k^2/M_s^2)^{2n}}\rightarrow 0$ for $k^2/M_s^2\rightarrow \infty$. Moreover, the asymptotic behavior is also the same for both t- and s-channels, as the even power, $2n,$ does not distinguish between space-like and time-like momentum exchange: $(\pm k^2)^{2n}=(k^2)^{2n}.$

Thus, we have shown that in both cases of bare and dressed propagator, and s- and t-channels, the asymptotic behaviors of the scattering amplitudes are the same for even powers ($2n$) of the d'Alembertian.

\subsubsection{Odd powers $\Box^{2n+1} $}

We now want to address the case of odd powers, $e^{-(k^2/M_s^2)^{2n+1}}$. Let us distinguish the cases of t-channel and s-channel.

\begin{itemize}
	
	\item {\bf $t$-channel scattering}: In the case of t-channel scattering, the momentum exchange is space-like, $k^2>0,$ and the asymptotic behavior for high energies turns out to be the same for both bare and dressed propagator, as it also happens for the case of even power. Indeed, for space-like momentum exchange, for $k^2/M_s^2\rightarrow \infty$, we have:
	\begin{equation}
	\mathcal{M}_{2n+1}\sim e^{-(k^2/M_s^2)^{2n+1}},
	\end{equation}
	as $\Sigma_{2n+1}(k)e^{-(k^2/M_s^2)^{2n+1}}\rightarrow 0$.

	\item  {\bf $s$-channel scattering}: In the case of s-channel scattering, the momentum exchange is time-like, $k^2<0,$ and it is clear that for high energy the propagator blows up. However, as we have already emphasized, non-locality is inherently off-shell, and as such quantum effects are not negligible and what is physically meaningful is the dressed propagator in the region of non-locality, $k^2>M_s^2.$ 
	
	By dressing the propagator, it so happens that for high energy the amplitude has the same asymptotic behavior as for the t-channel, with the same exponential suppression, as one would expect for consistency. We will show this by making an explicit calculation for the simplest case $n=1$. In this case the s-channel amplitude with dressed propagator is given by:
	\begin{equation}
	\mathcal{M}_{1}\sim  \lambda^2\frac{e^{-k^2/M_s^2}}{k^2+\Sigma_1(k)e^{-k^2/M_s^2}}.\label{s-chan}
	\end{equation}
	Note that the self-energy $\Sigma_1 (k)$ at $1$-loop can be explicitly computed by performing the integration in Euclidean space and then analytically continuing back to Minkowski the momentum $k$; indeed by using the Schwinger parameterization we obtain:
	\begin{equation}
	\begin{array}{rl}
	\Sigma_1(k)= & \displaystyle \frac{\lambda^2}{16\pi^2} \int\limits_{\frac{1}{M_s^2}}^{\infty}dt_1\int\limits_{\frac{1}{M_s^2}}^{\infty}dt_2 \frac{e^{-k^2\frac{t_1t_2}{t_1+t_2}}}{(t_1+t_2)^2}   \\
	=& \displaystyle \frac{\lambda^2}{16\pi^2}\left[\frac{2M_s^2}{k^2}\left(e^{-k^2/2M_s^2}-e^{-k^2/M_s^2}\right)+{\rm Ei}\left(-\frac{k^2}{2M_s^2}\right)-{\rm Ei}\left(-\frac{k^2}{M_s^2}\right)\right],\label{1-loop self-energy}
	\end{array}
	\end{equation}
	which in the high energy regime goes to zero for space-like momentum exchange, while diverges for time-like exchange. In particular, the asymptotic behavior of the self-energy for $k^2/M_s^2\rightarrow \infty$ is given by:
	\begin{equation}
	\Sigma_{1}(k)\sim e^{-k^2/M_s^2}.
	\end{equation}
	It is now clear that the interplay between the divergences of the self-energy and the bare propagator, for time-like exchange, is such that the fully dressed $s$-channel scattering amplitude turns out to be exponentially suppressed in the UV regime, showing the same behavior as in the case of the $t$-channel. Indeed, the asymptotic behavior of the $s$-channel amplitude in Eq.\eqref{s-chan} is given by\footnote{Note that such a result relies on the fact that we can define the dressed propagator by summing up the geometric series in Eq.\eqref{5.21}. However, the geometric series in Eq.\eqref{5.21} converges if and only if $|\Pi(k)\Sigma(k)|<1,$ and such a condition is always satisfied if we define the series in the Euclidean signature where the momenta are space-like.}
	\begin{equation}
	\mathcal{M}_{1}\sim  \lambda^2\frac{e^{-k^2/M_s^2}}{k^2+\lambda^2 e^{-k^2/M_s^2}\cdot e^{-k^2/M_s^2}}\sim e^{k^2/M_s^2},
	\end{equation}
	which is exponentially suppressed in the UV regime, $k^2/M_s^2\rightarrow \infty$,  with $k^2<0.$
	
\end{itemize}

Therefore, we have shown that all scattering amplitudes in $\phi^3$-theory, for any kind of channel,\footnote{Note that the case of $u$-channel is similar to the $t$-channel where the momentum exchange is space-like.} and for any power $n$ of the d'Alembertian, are physically well-defined and have exactly the same UV behavior, once the propagator is consistently dressed. However, a more general study is needed to include more complicated theories and understand whether the dressing procedure can always help to cure the s-channel divergence for form-factors with odd powers of the d'Alembertian.

Furthermore, in the case of $1$-loop or multi-loops amplitudes, we still have finite results, for example, quartic interaction with propagators that are exponentially suppressed was considered in Ref.\cite{Biswas:2014yia}, where scattering amplitudes $(\phi\phi\rightarrow\phi\phi)$ and cross sections were computed. One can also consider loop-amplitudes for decay of unstable particles $(\phi\rightarrow\psi\psi)$, but in this case, although there is no UV divergence, the amplitudes blow up for some values of the integration variables. However, such divergent integrals can be regularized with appropriate prescriptions and are still physically meaningful; see for example Refs.\cite{witten-epsilon,sen-epsilon}. 



\subsection{Dressed propagator and unitarity}\label{dress-unitar}

The concept of dressed propagator has been very useful to obtain a singularity-free Euclidean correlator and to make the amplitudes well defined. In this subsection we wish to analyze the pole structure of the dressed propagator in infinite derivative scalar field theory; as an example we will consider a massless scalar field in $\phi^3$-theory, with the simplest choice $f(\Box)=\Box/M_s^2.$ 

\subsubsection{Infinite massive complex conjugate poles} 

We already know that the infinite derivative field theory under study is unitary at the tree-level: the propagator has a pole at $k^2=0$ with positive residue corresponding to one single physical scalar degree of freedom. In the case of the dressed propagator, the pole structure is described by the following equation (see Eqs.(\ref{5.21},\ref{s-chan})):
\begin{equation}
k^2+\Sigma(k)e^{-k^2/M_s^2}=0.
\end{equation}
As a first example, let us consider the simpler complex equation
\begin{equation}
z=ce^{z},\label{complex-eq}
\end{equation}
where $-k^2/M_s^2=:z=x+iy$ and $c$ is a finite positive constant. We can now study the complex equation in Eq.\eqref{complex-eq} and understand how many and which kind of solutions it has.
\begin{itemize}
	
	\item First all note that Eq.\eqref{complex-eq} can have at most two real solutions, when the line $z=x$ intersects the exponential $ce^{z}=ce^{x}.$ We can check that one of the solution can be associated to the usual shifted mass value, while the second one comes with a negative residue, whose mass value is larger than the scale of non-locality, $(x=-k^{2}/M_{s}^2> 1)$. Moreover, the smaller is $c,$ which is related to the coupling constant, the larger will be the mass of this ghost-mode beyond the scale of non-locality\footnote{See Ref.\cite{Hashi:2018kag} for a similar real particle spectrum beyond the scale of non-locality but in the context of Higgs mechanism.}. From a physical point of view, as we have stressed in the previous section, such a pole can not correspond to any physical state: beyond the physical cut-off $M_s$, there are no asymptotic states which can be constructed, and no physical measurements can be made for $k^{2}>M_{s}^{2}$.
	
	\item Secondly, we can check whether there are any massive complex poles. By decomposing $z$ in real and imaginary components, we obtain two equations coming from the real and imaginary parts of Eq.\eqref{complex-eq}:
	\begin{equation}
	\begin{array}{rl}
	x=&-ce^{x}{\rm cos}y,\\
	y=&-ce^{x}{\rm sin}y,
	\end{array}
	\end{equation} 
	which can be rewritten as
	\begin{equation}
	x=\frac{y}{{\rm tan}y},\,\,\,\,\,\,\,y=-ce^{y/{\rm tan}y}{\rm sin}y.\label{system-compl-eqs}
	\end{equation} 
	We note that this system of two equations has an infinite number of solutions, which means that the dressed propagator has an {\it infinite number of massive complex poles}. Moreover, by studying the system of equations in Eq.\eqref{system-compl-eqs} we can easily understand that if $x+iy$ is a solution, also its complex conjugate, $x-iy,$ will be a solution as the equations are unchanged under the transformation $y\rightarrow-y.$
\end{itemize}


\subsubsection{Unitarity with infinite complex conjugate poles}

A quantum field theory is unitarity if and only if the S-matrix is unitary:
\begin{equation}
S^{\dagger}S=\mathbb{I},
\end{equation}
which, by introducing the amplitude $T$ through $S=\mathbb{I}+iT,$ can be also expressed as
\begin{equation}
2{\rm Im}\left\lbrace T\right\rbrace=T^{\dagger}T>0. 
\end{equation}
From the last equation, we can immediately see that the imaginary part of an amplitude $T$ has to be always positive; as an example, we can consider an amplitude $\phi\phi\rightarrow\phi\phi$ in $\phi^3$-theory. We now wish to point out that the presence of extra poles in the dressed propagator is harmless, as far as perturbative unitarity is concerned \cite{anselmi}. Indeed, in quantum field theory what is needed to prove perturbative unitarity is the tree-level propagator and higher loop amplitudes constructed in terms of tree-level propagators\footnote{As a clear example we can consider quantum electrodynamics, which is unitary at the perturbative level. At the same time, it also known that by dressing the propagator an extra ghost-like pole emerges, which is related to the Landau pole. However, such a more complicated pole structure of the dressed photon propagator does not spoil perturbative unitarity.}. However, it is interesting to understand which is the pole structure of the dressed propagator and whether the presence of infinite massive complex poles may create some instabilities.

If we consider a {\it tree} level internal propagator we already know that the amplitude preserves unitarity, but now we can ask what happens if we have infinite conjugate complex poles.\footnote{The amplitude $T$ for a tree level internal propagator, $\Pi(k)$, is given by:
	\begin{equation}
	T=\lambda^2 \Pi(k)=\lambda^2 \frac{e^{-k^2/M_s^2}}{k^2-i\epsilon}.
	\end{equation}
	If we compute the imaginary part of the amplitude in the last equation, we obtain:
	\begin{equation}
	{\rm Im}\left\lbrace T\right\rbrace 
	=\displaystyle \lambda^2 \pi \delta(k^2)
	\equiv \displaystyle \lambda^2 {\rm Im}\left\lbrace {\rm Res}\left(\Pi(k)\right)\right\rbrace_{k^2=0},
	\end{equation}
	which is positive if and only if the residue of the propagator is positive. In the simple case of a scalar field we obtain: ${\rm Im}\left\lbrace {\rm Res}\left(\Pi(k)\right)\right\rbrace_{k^2=0}=\pi \delta(k^2)>0,$ i.e. the unitarity condition is preserved.}
We will show that the imaginary part of the amplitude can be non-negative also in the case of the dressed propagator. The amplitude with an internal dressed propagator in Minkowski space reads:
\begin{equation}
T=\lambda^2 \mathcal{G}(k)=\lambda^2 \frac{e^{-k^2/M_s^2}}{k^2-i\Sigma(k)e^{-k^2/M_s^2}},\label{dress-propag}
\end{equation}
$\Gamma\sim \Sigma(k)e^{-k^2/M_s^2}$ represents the width of the particle. If we now compute the imaginary part of Eq.\eqref{dress-propag} we obtain:
\begin{equation}
{\rm Im}\left\lbrace \lambda^2\mathcal{G}(k)\right\rbrace =\lambda^2{\rm Im}\left\lbrace\frac{e^{-k^2/M_s^2}}{k^2-i\Sigma(k)e^{-k^2/M_s^2}}\right\rbrace=\lambda^2\frac{e^{-2k^2/M_s^2}\Sigma(k)}{k^4+\Sigma^2(k)e^{-2k^2/M_s^2}}. \label{dress-prop-poles}
\end{equation}
Note that the sign in Eq.\eqref{dress-prop-poles} is determined by the sign of $\Sigma(k).$ First of all, we can observe that for $\phi^4$-theory the self-energy at $1$-loop is a positive constant, see the subsection \ref{DP}, so that the imaginary part of the dressed propagator is positive too. Furthermore, we can also perform the same check for a non-constant self-energy $(\Sigma(k)\neq c)$. For instance in the case of $\phi^3$-theory we need to consider the $1$-loop expression in Eq.\eqref{1-loop self-energy} which can be checked to be always positive for any value of $k.$

Therefore, we have shown that also for a dressed propagator the imaginary part can be still positive consistently with the optical theory, despite the presence of extra poles.

\section{Summary and conclusions}\label{conlus}

In this paper we have studied quantum aspects of infinite derivative scalar field theory. We have shown that the action can be made non-local by introducing Lorentz invariant analytic form factor either in the kinetic operator or in the interaction vertex. We have shown that in order to not introduce any {\it ghost-like} degree of freedom, we require the form factors ought to be {\it exponential of entire function}; in particular, we have considered exponentials of polynomials of the d'Alembertian $\Box$, see Eq.\eqref{eq:14}. 

We have explicitly shown that the non-local propagator is not simply defined in terms of a time-ordered product, unlike the local theory, but it is made of an acausal contribution. Moreover, the non-local analog of the retarded Green function assumes an acausal behavior, indeed it is non-vanishing for space-like separations. As a consequence, also the local commutativity condition is violated in presence of non-local interaction.

In the non-local region, we cannot define any concept of space and time due to the presence of acausal effects. Such a statement is also mathematically justified by the fact that amplitudes are ill-defined with the Minkowski signature due to the presence of the exponential form factors, which can diverge along some direction in the complex plane, making it impossible to define the Wick-rotation. For this reasons, the recipe is to define the theory in the Euclidean space, where all the amplitudes can be well-defined, and after having performed the computations, we can analytically continue back the external momenta to the Minkowski signature. We have studied the structure of the Euclidean $2$-point correlation function, and shown that it is non-singular at the Euclidean origin.

We have discussed that scattering amplitudes for momenta in the regime $k^2\geq M_s^2,$ where the vertices are smeared out in momentum space. Indeed, non-locality and acausality manifest as off-shell phenomena, which means that all amplitudes have to be seen as quantum for momenta $k^2\geq M_s^2,$ and all perturbative quantum corrections have to be taken into account by dressing propagators and vertices. In this way all scattering amplitudes, for any channel, and for both odd and even powers of the d'Alembertian, turn out to be well-defined. We have found that in the dressed propagator, besides the physical mass shift,  there is the presence of one real massive ghost, and an infinite number of complex poles. However, these extra poles do not spoil perturbative unitarity.

Although, infinite derivative quantum field theory shows many interesting features, there are still some open questions that need to be possibly answered. For instance, the usual K\"allén-Lehmann representation for the non-local propagator is not possible; see Ref.\cite{Tomboulis:1997gg} for some attempts aimed to generalize such a representation for the propagator to the case of infinite derivative interactions. Furthermore, systematic methods to proof the unitarity and macrocausality~\footnote{Macrocausality is a generalization of the concept of causality in which one can have the presence of acausal effects at microscopic scales $(t,r\leq1/M_s)$, but physics is still causal on macroscopic scales $(t,r\gg1/M_s)$.} conditions at the level of the S-matrix have not been developed yet. In the local quantum field theory, it is well known that the unitarity can be proven by using the {\it largest time equation} \cite{anselmi}. Such an approach strongly relies on two crucial hypothesis: (i) the propagator has a time-ordered structure, (ii) vertices are local. It is clear that when the principle of locality is given up at the level of interaction, and infinite derivative are introduced, the largest time equation cannot be consistently applied to check the unitarity, as the propagator is not simply a time ordered product and the interaction vertices become non-local.

The presence of non-local interaction seems to be very important to avoid singularities of several types, thanks to its regularizing nature. Infinite derivative field theory might be very important to construct a consistent theory of quantum gravity, especially when dealing with blackhole physics; see for example Ref.\cite{Koshelev:2017bxd,Buoninfante:2018rlq}. For these reasons, we strongly believe that infinite derivative field theories deserve further and deeper investigations.

\section*{Acknowledgement}
The authors would like to thank Terry Tomboulis, Elisabetta Pallante and Valery Frolov for helpful and insightful discussions. AM’s research is financially supported by Netherlands Organisation for Scientific Research (NWO) grant number 680-91-119.

\section{Appendix}

\subsection{Principal value computation for acausal Green function}\label{app-princ}

We now want to show the computation that leads to the expression in Eq.\eqref{77} for the acausal Green function in infinite derivative field theory. In particular, we want to compute the principal-value integral in Eq. \eqref{70} that we recall for clarity:
\begin{equation}
I_{\scriptscriptstyle PV}=\frac{1}{16i\pi^3}\frac{1}{r} \int\limits_{-\infty}^{\infty} kdk \,\,{\rm P.V.} \!\int\limits_{-\infty}^{\infty}dk^{0}\frac{e^{-\left(\frac{-k_0^2+k^2}{M_s^{2}}\right)^{2n}}}{k_0^2-k^2}\left(e^{i(kr-k^0t)}-e^{-i(kr+k^0t)}\right), \label{70.2}
\end{equation}
where, let us remind that $k\equiv |\vec{k}|$ and $\omega_{\vec{k}}=k$, as we are working with the massless case.

Since we are interested in the  modification of the local retarded Green function we will consider the case $t>0$\footnote{If we considered the case $t<0$ we would study the  modification of the  advanced Green function.}. To compute $I_{\scriptscriptstyle PV}$ we need to consider several cases corresponding to different regions of the planes $t$-$r$ and $k^0$-$k$. As for the plane $t$-$r$ we have to distinguish~\footnote{In Ref.\cite{pais} the authors consider the same calculation for the case $2n=2$.}:
\begin{enumerate}
	\item $t>0,\,\, t^2>r^2 \Longleftrightarrow (x-y)^2<0$ (time-like separation):
	\begin{equation*}
	t=\rho^{1/2}{\rm cosh}^2\alpha,\,\,\,r=\rho^{1/2}{\rm sinh}^2\alpha,\,\,\,t^2-r^2=\rho>0;
	\end{equation*}
	\item $t>0,\,\, t^2<r^2 \Longleftrightarrow (x-y)^2>0$ (space-like separation):
	\begin{equation*}
	t=\rho^{1/2}{\rm cosh}^2\alpha,\,\,\,r=\rho^{1/2}{\rm sinh}^2\alpha,\,\,\,t^2-r^2=\rho<0.
	\end{equation*}
\end{enumerate}
Instead, as for the plane $k^0$-$k$ we will split the double integral in Eq. \eqref{70.2} in the following two regions:
\begin{enumerate}[label=(\roman*)]
	\item $k_0^2>k^2$:
	\begin{equation*}
	k=R{\rm sinh}\beta,\,\,\,k^0=R{\rm cosh}\beta,\,\,\,R^2=k_0^2-k^2>0,\,\,\,-\infty<\beta,R<\infty;
	\end{equation*}

	\item $k_0^2<k^2$:
	\begin{equation*}
	k=R{\rm cosh}\beta,\,\,\,k^0=R{\rm sinh}\beta,\,\,\,-R^2=k_0^2-k^2<0,\,\,\,-\infty<\beta,R<\infty.
	\end{equation*}
\end{enumerate}
By moving to the new integration variables $R,\beta$ we get a Jacobian factor $|R|$ so that the integral in Eq.\eqref{70.2} in the case 1. ($\rho>0$) reads
\begin{equation}
\!\begin{array}{ll}
\displaystyle I_{\scriptscriptstyle PV}= \displaystyle \frac{i}{16\pi^3 r}\int\limits_{-\infty}^{\infty}dRd\beta e^{-R^{4n}/M_s^{4n}}\frac{|R|}{R}\left\lbrace {\rm sinh}\beta\left[e^{-iR\rho^{1/2}{\rm cosh}(\beta -\alpha)}- e^{-iR\rho^{1/2}{\rm cosh}(\beta+\alpha)}\right]\right. &\\
\,\,\,\,\,\,\,\,\,\,\,\,\,\,\,\,\,\,\,\,\,\,\,\,\,\,\,\,\,\,\,-\displaystyle \left.{\rm cosh}\beta\left[e^{iR\rho^{1/2}{\rm sinh}(\beta -\alpha)}- e^{-iR\rho^{1/2}{\rm sinh}(\beta+\alpha)}\right]\right\rbrace  &\\
\,\,\,\,\,\,\,\,\,\,\,= \displaystyle -\frac{1}{8\pi^3 r}\!\!\int\limits_{-\infty}^{\infty}\!\!\!dRd\beta e^{-R^{4n}/M_s^{4n}}\frac{|R|}{R}\left\lbrace {\rm sinh}\beta {\rm sin}\left(R\rho^{1/2}{\rm cosh}(\beta+\alpha)\right)\right.\\
\,\,\,\,\,\,\,\,\,\,\,\,\,\,\,\,\,\,\,\,\,\,\,\,\,\,\,\,\,\,\displaystyle \left.-{\rm cosh}\beta{\rm sin}\left(R\rho^{1/2}{\rm sinh}(\beta+\alpha)\right)\right\rbrace. &
\end{array}\label{71}
\end{equation}
Defining the new integration variable $\theta=\beta +\alpha$, the integral in Eq.\eqref{71} becomes:
\begin{equation}
\begin{array}{rl}
\displaystyle I_{\scriptscriptstyle PV}= & \displaystyle \frac{1}{8\pi^3 \rho^{1/2}}\int\limits_{-\infty}^{\infty}dRd\theta e^{-R^{4n}/M_s^{4n}}\frac{|R|}{R}\left\lbrace {\rm cosh}\theta {\rm sin}\left(R\rho^{1/2}{\rm cosh}\theta\right)-{\rm sinh}\theta{\rm sin}\left(R\rho^{1/2}{\rm sinh}\theta\right)\right\rbrace\\
= & -\displaystyle \frac{1}{\pi^3}\frac{\partial}{\partial\rho}\int\limits_{0}^{\infty}\frac{dR}{R} e^{-R^{4n}/M_s^{4n}}\int\limits_{0}^{\infty}d\theta\left\lbrace {\rm cos}\left(R\rho^{1/2}{\rm cosh}\theta\right)-{\rm cos}\left(R\rho^{1/2}{\rm sinh}\theta\right)\right\rbrace.
\end{array}\label{72}
\end{equation}
The integrals in $\theta$-variable can be expressed in terms of Bessel functions \cite{watson}:
\begin{equation}
\int\limits_{0}^{\infty}d\theta{\rm cos}\left(R\rho^{1/2}{\rm sinh}\theta\right)=K_0(R\rho^{1/2}),\,\,\,\,\int\limits_{0}^{\infty}d\theta{\rm cos}\left(R\rho^{1/2}{\rm cosh}\theta\right)=-\frac{\pi}{2}Y_0(R\rho^{1/2});\label{73}
\end{equation}
then, by introducing the dimensionless variable $\zeta=R\rho^{1/2}$, the principal-value integral in Eq.\eqref{72} becomes
\begin{equation}
I_{\scriptscriptstyle PV}=\frac{1}{\pi^3}\frac{\partial}{\partial\rho}\int\limits_{0}^{\infty}\frac{d\zeta}{\zeta}e^{-\frac{\zeta^{4n}}{M_s^{4n}\rho^{2n}}}\left[K_0(\zeta)+\frac{\pi}{2}Y_0(\zeta)\right].
\label{75}
\end{equation}
The last result holds for the case 1. when $\rho>0$, but we can also take into account the case 2., when $\rho<0$, by considering the following expression:
\begin{equation}
I_{\scriptscriptstyle PV}=\frac{1}{\pi^3}\frac{\partial}{\partial\rho}\left\lbrace \varepsilon(\rho) \int\limits_{0}^{\infty}\frac{d\zeta}{\zeta}e^{-\frac{\zeta^{4n}}{M_s^{4n}\rho^{2n}}}\left[K_0(\zeta)+\frac{\pi}{2}Y_0(\zeta)\right]\right\rbrace,
\label{76}
\end{equation}
where the function $\varepsilon(\rho)$ is equal to $+1$ if $\rho>0$ (time-like separation), while it is $-1$ if $\rho<0$ (space-like separation). The result in Eq.\eqref{76} corresponds to the integral in Eq.\eqref{71.2}.

The integral in Eq.\eqref{76} can be computed analytically for $2n=2$ and can be expressed in terms of the Meijer-G functions \cite{luke}; indeed for space-like separation $(\rho<0)$ one has
\begin{equation}
\begin{array}{rl}
I_{\scriptscriptstyle PV}= & \displaystyle \frac{1}{\pi^3}\frac{\partial}{\partial\rho}\left\lbrace \int\limits_{0}^{\infty}\frac{d\zeta}{\zeta}e^{-\frac{\zeta^{4}}{M_s^{4}\rho^{2}}}\left[K_0(\zeta)+\frac{\pi}{2}Y_0(\zeta)\right]\right\rbrace\\
= & \displaystyle \frac{2}{\pi^3M_s^4\rho^3}\left\lbrace \int\limits_{0}^{\infty}d\zeta e^{-\frac{\zeta^{4}}{M_s^{4}\rho^{2}}}\zeta^3\left[K_0(\zeta)+\frac{\pi}{2}Y_0(\zeta)\right]\right\rbrace\\
= & \displaystyle \frac{1}{2\pi^4}\frac{1}{\rho}\left\lbrace G_{2,5}^{4,1}\left(\begin{array}{l}
0\\
0,0,\frac{1}{2},\frac{1}{2}
\end{array}\Biggr|\frac{M_{s}^{4}\rho^{2}}{256}\right)+2\pi^{2}G_{3,6}^{4,1}\left(\begin{array}{l}
0,-\frac{1}{4},\frac{1}{4}\\
0,0,\frac{1}{2},\frac{1}{2},-\frac{1}{4},\frac{1}{4}
\end{array}\Biggr|\frac{M_{s}^{4}\rho^{2}}{256}\right)\right\rbrace,
\end{array}
\label{78}
\end{equation}
which explains the expression in Eq.\eqref{77} for the acausal retarded Green function $G_R.$

\end{document}